\documentclass[twocolumn]{aastex62}

\usepackage{float}
\usepackage{graphicx}
\usepackage{xcolor}
\usepackage{amsmath}

\shortauthors{Kanodia et al. 2020.}
\shorttitle{A warm super Neptune orbiting an M dwarf}

\received{20th May 2020}
\revised{24th June 2020}
\accepted{25th June 2020}
\submitjournal{ApJ}

\newcommand\tess{\textit{TESS}}
\newcommand\gaia{\textit{Gaia}}
\newcommand\kms{$\textrm{km/s}$}
\newcommand\ms{$\textrm{m/s}$}
\newcommand\cms{$\textrm{cm/s}$}

\newcommand\teff{T$_{\rm{eff}}$}

\newcommand{\unit}[1]{\ensuremath{\, \mathrm{#1}}} 

\begin{document}
\title{TOI-1728b: The Habitable-zone Planet Finder confirms a warm super Neptune orbiting an M dwarf host}

\author[0000-0001-8401-4300]{Shubham Kanodia}
\affil{Department of Astronomy \& Astrophysics, 525 Davey Laboratory, The Pennsylvania State University, University Park, PA, 16802, USA}
\affil{Center for Exoplanets and Habitable Worlds, 525 Davey Laboratory, The Pennsylvania State University, University Park, PA, 16802, USA}

\author[0000-0003-4835-0619]{Caleb I. Ca\~nas}
\affil{Department of Astronomy \& Astrophysics, 525 Davey Laboratory, The Pennsylvania State University, University Park, PA, 16802, USA}
\affil{Center for Exoplanets and Habitable Worlds, 525 Davey Laboratory, The Pennsylvania State University, University Park, PA, 16802, USA}
\affiliation{NASA Earth and Space Science Fellow}

\author[0000-0001-7409-5688]{Gudmundur Stefansson}
\affiliation{Henry Norris Russell Fellow}
\affiliation{Department of Astrophysical Sciences, Princeton University, 4 Ivy Lane, Princeton, NJ 08540, USA}

\author[0000-0001-6160-5888]{Joe P.\ Ninan}
\affil{Department of Astronomy \& Astrophysics, 525 Davey Laboratory, The Pennsylvania State University, University Park, PA, 16802, USA}
\affil{Center for Exoplanets and Habitable Worlds, 525 Davey Laboratory, The Pennsylvania State University, University Park, PA, 16802, USA}

\author[0000-0003-1263-8637]{Leslie Hebb}
\affiliation{Department of Physics, Hobart and William Smith Colleges, 300 Pulteney Street, Geneva,
NY, 14456, USA}

\author[0000-0002-9082-6337]{Andrea S.J. Lin}
\affil{Department of Astronomy \& Astrophysics, 525 Davey Laboratory, The Pennsylvania State University, University Park, PA, 16802, USA}
\affil{Center for Exoplanets and Habitable Worlds, 525 Davey Laboratory, The Pennsylvania State University, University Park, PA, 16802, USA}

\author[0000-0002-0934-3574]{Helen Baran}
\affil{Department of Astronomy \& Astrophysics, 525 Davey Laboratory, The Pennsylvania State University, University Park, PA, 16802, USA}
\affil{Center for Exoplanets and Habitable Worlds, 525 Davey Laboratory, The Pennsylvania State University, University Park, PA, 16802, USA}

\author[0000-0001-8222-9586]{Marissa Maney}
\affil{Department of Astronomy \& Astrophysics, 525 Davey Laboratory, The Pennsylvania State University, University Park, PA, 16802, USA}
\affil{Center for Exoplanets and Habitable Worlds, 525 Davey Laboratory, The Pennsylvania State University, University Park, PA, 16802, USA}

\author[0000-0002-4788-8858]{Ryan C. Terrien}
\affil{Department of Physics and Astronomy, Carleton College, One North College Street, Northfield, MN 55057, USA}

\author[0000-0001-9596-7983]{Suvrath Mahadevan}
\affil{Department of Astronomy \& Astrophysics, 525 Davey Laboratory, The Pennsylvania State University, University Park, PA, 16802, USA}
\affil{Center for Exoplanets and Habitable Worlds, 525 Davey Laboratory, The Pennsylvania State University, University Park, PA, 16802, USA}

\author[0000-0001-9662-3496]{William D. Cochran}
\affil{McDonald Observatory and Department of Astronomy, The University of Texas at Austin}
\affil{Center for Planetary Systems Habitability, The University of Texas at Austin}

\author[0000-0002-7714-6310]{Michael Endl}
\affil{McDonald Observatory and Department of Astronomy, The University of Texas at Austin}
\affil{Center for Planetary Systems Habitability, The University of Texas at Austin}

\author[0000-0002-3610-6953]{Jiayin Dong}
\affil{Department of Astronomy \& Astrophysics, 525 Davey Laboratory, The Pennsylvania State University, University Park, PA, 16802, USA}
\affil{Center for Exoplanets and Habitable Worlds, 525 Davey Laboratory, The Pennsylvania State University, University Park, PA, 16802, USA}

\author[0000-0003-4384-7220]{Chad F.\ Bender}
\affil{Steward Observatory, The University of Arizona, 933 N.\ Cherry Ave, Tucson, AZ 85721, USA}

\author[0000-0002-2144-0764]{Scott A. Diddams}
\affil{Time and Frequency Division, National Institute of Standards and Technology, 325 Broadway, Boulder, CO 80305, USA}
\affil{Department of Physics, University of Colorado, 2000 Colorado Avenue, Boulder, CO 80309, USA}

\author[0000-0001-6545-639X]{Eric B.\ Ford}
\affil{Department of Astronomy \& Astrophysics, 525 Davey Laboratory, The Pennsylvania State University, University Park, PA, 16802, USA}
\affil{Center for Exoplanets and Habitable Worlds, 525 Davey Laboratory, The Pennsylvania State University, University Park, PA, 16802, USA}
\affil{Institute for CyberScience, The Pennsylvania State University, University Park, PA, 16802, USA}

\author[0000-0002-0560-1433]{Connor Fredrick}
\affil{Time and Frequency Division, National Institute of Standards and Technology, 325 Broadway, Boulder, CO 80305, USA}
\affil{Department of Physics, University of Colorado, 2000 Colorado Avenue, Boulder, CO 80309, USA}

\author[0000-0003-1312-9391]{Samuel Halverson}
\affil{Jet Propulsion Laboratory, 4800 Oak Grove Drive, Pasadena, CA 91109, USA}

\author[0000-0002-1664-3102]{Fred Hearty}
\affil{Department of Astronomy \& Astrophysics, 525 Davey Laboratory, The Pennsylvania State University, University Park, PA, 16802, USA}
\affil{Center for Exoplanets and Habitable Worlds, 525 Davey Laboratory, The Pennsylvania State University, University Park, PA, 16802, USA}

\author[0000-0001-5000-1018]{Andrew J. Metcalf}
\affiliation{Space Vehicles Directorate, Air Force Research Laboratory, 3550 Aberdeen Ave. SE, Kirtland AFB, NM 87117, USA}
\affiliation{Time and Frequency Division, National Institute of Technology, 325 Broadway, Boulder, CO 80305, USA} 
\affiliation{Department of Physics, 390 UCB, University of Colorado Boulder, Boulder, CO 80309, USA}

\author[0000-0002-0048-2586]{Andrew Monson}
\affil{Department of Astronomy \& Astrophysics, 525 Davey Laboratory, The Pennsylvania State University, University Park, PA, 16802, USA}
\affil{Center for Exoplanets and Habitable Worlds, 525 Davey Laboratory, The Pennsylvania State University, University Park, PA, 16802, USA}

\author[0000-0002-4289-7958]{Lawrence W. Ramsey}
\affil{Department of Astronomy \& Astrophysics, 525 Davey Laboratory, The Pennsylvania State University, University Park, PA, 16802, USA}
\affil{Center for Exoplanets and Habitable Worlds, 525 Davey Laboratory, The Pennsylvania State University, University Park, PA, 16802, USA}

\author[0000-0003-0149-9678]{Paul Robertson}
\affil{Department of Physics and Astronomy, The University of California, Irvine, Irvine, CA 92697, USA}

\author[0000-0001-8127-5775]{Arpita Roy}
\affiliation{Robert A. Millikan Postdoctoral Fellow}
\affil{Department of Astronomy, California Institute of Technology, Pasadena, CA 91125, USA}

\author[0000-0002-4046-987X]{Christian Schwab}
\affil{Department of Physics and Astronomy, Macquarie University, Balaclava Road, North Ryde, NSW 2109, Australia}

\author[0000-0001-6160-5888]{Jason T.\ Wright}
\affil{Department of Astronomy \& Astrophysics, 525 Davey Laboratory, The Pennsylvania State University, University Park, PA, 16802, USA}
\affil{Center for Exoplanets and Habitable Worlds, 525 Davey Laboratory, The Pennsylvania State University, University Park, PA, 16802, USA}


\correspondingauthor{Shubham Kanodia}
\email{shbhuk@gmail.com}

\begin{abstract}
We confirm the planetary nature of TOI-1728b using a combination of ground-based photometry, near-infrared Doppler velocimetry and spectroscopy with the Habitable-zone Planet Finder. TOI-1728 is an old, inactive M0 star with \teff{} $= 3980^{+31}_{-32}$ K, which hosts a transiting super Neptune at an orbital period  of $\sim$ 3.49 days. Joint fitting of the radial velocities and TESS and ground-based transits yields a planetary radius of $5.05_{-0.17}^{+0.16}$ R$_{\oplus}$, mass $26.78_{-5.13}^{+5.43}$ M$_{\oplus}$ and eccentricity $0.057_{-0.039}^{+0.054}$. We estimate the stellar properties, and perform a search for He 10830 \AA  ~absorption during the transit of this planet and claim a null detection with an upper limit of 1.1$\%$ with 90\% confidence. A deeper level of He 10830 \AA ~ absorption has been detected in the planet atmosphere of GJ 3470b, a comparable gaseous planet. TOI-1728b is the largest super Neptune —the intermediate subclass of planets between Neptune and the more massive gas-giant planets—discovered around an M dwarf. With its relatively large mass and radius, TOI-1728 represents a valuable datapoint in the M-dwarf exoplanet mass-radius diagram, bridging the gap between the lighter Neptune-sized planets and the heavier Jovian planets known to orbit M-dwarfs. With a low bulk density of $1.14_{-0.24}^{+0.26}$ g/cm$^3$, and orbiting a bright host star (J $\sim 9.6$, V $\sim 12.4$), TOI-1728b is also a promising candidate for transmission spectroscopy both from the ground and from space, which can be used to constrain planet formation and evolutionary models.
\end{abstract}


\section{Introduction} \label{sec:intro}

\begin{figure*}[!t] 
\centering
\includegraphics[width=0.98\textwidth]
{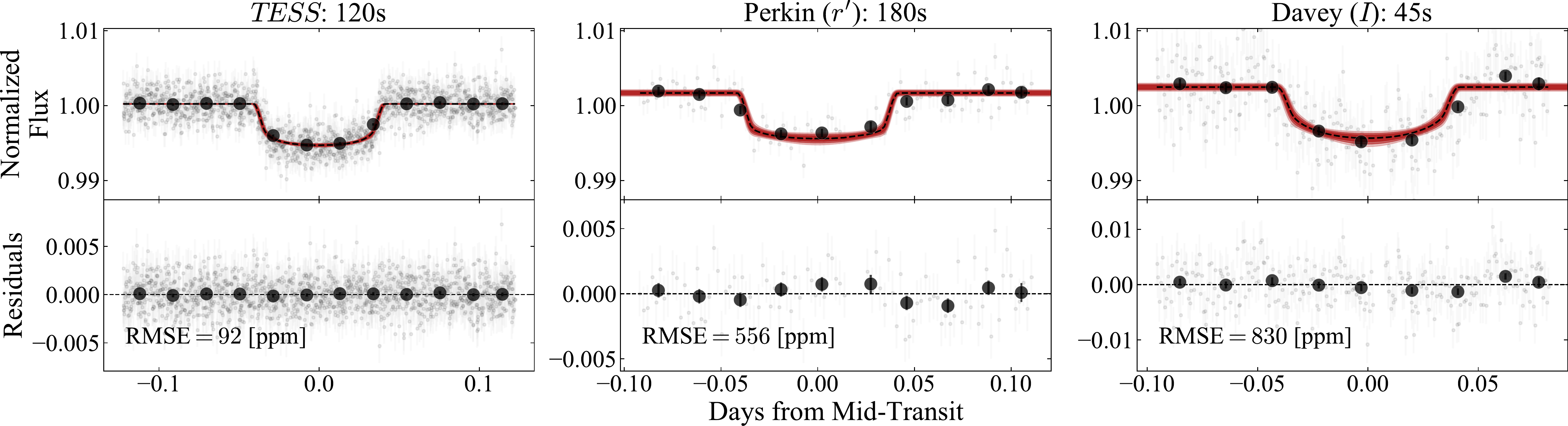}
\caption{Composite panel showing the unbinned and binned transit observations for TOI-1728.  In each case, the best-fitting model is plotted as a dashed line while the shaded regions denote the \(1\sigma\) (darkest), \(2\sigma\), and \(3\sigma\) range of the derived posterior solution. The exposure time for each of the instruments along with the filter used for the observation is given along with the transit light curve. The residuals show the RMS error calculated for the binned data  (30 min bins) for each instrument. \textbf{Left}: \tess{} phase folded light curve from Sector 20. \textbf{Middle:} Ground-based transit using the Perkin telescope and defocussed observations. \textbf{Right:} Diffuser assisted photometry using the Engineered Diffuser on the CDK 0.6 m telescope at Davey laboratory.} \label{fig:photometry}
\end{figure*}

Neptune-sized exoplanets ($2 R_{\oplus} < R_p < 6 R_{\oplus}$) represent not only a fairly common population around stars \citep[$> 25\%$,][]{2014Natur.509..593B, 2019AJ....158..109H}, but also a transitional population between rocky terrestrial planets and Jupiter-like gas giants. Of these, transiting super Neptunes ($17 M_{\oplus} < M_p < 57 M_{\oplus}$; \cite{2015ApJ...813..111B}) with mass measurements are important in trying to understand theories of planet formation \citep{Crossfield2017}. Constraints on their chemical abundances (O/H, C/H and C/O) can help inform theories of planet formation and migration, i.e., whether the planets formed in-situ or formed farther away beyond the ice lines and migrated inwards \citep{2017MNRAS.469.4102M}. Furthermore, the atmospheric composition obtained by transmission spectroscopy can be used to understand the protoplanetary disk chemistry since it is expected that planetesimal accretion forms the main source of heavy elements in their atmospheres  \citep{2016ApJ...832...41M}. A subset of these exo-Neptunes with equilibrium temperatures of $\sim 800$ -- $1200$ K are referred to as ``warm Neptunes", and are expected to exhibit a wide diversity in the atmospheric elemental abundances, as well as atmospheres which are dominated by CO instead of CH$_4$ \citep{2020arXiv200410106G}.

M dwarfs, the most common stars in the Galaxy \citep{1997AJ....113.2246R, henry_solar_2006}, represent lucrative targets for exoplanet transmission spectroscopy due to their large planet-to-star radii ratios \citep{Batalha2017}.  Compared to planets around earlier type host stars, the lower luminosity results in smaller semi-major axes for comparable insolation fluxes. In addition, the lower stellar masses (in comparison with solar-type stars), amplifies the radial velocity (RV) signal amplitude for planets orbiting M dwarfs as opposed to solar type stars, for comparable insolation flux.

 Using RVs from the near infrared (NIR) Habitable zone Planet Finder (HPF) spectrometer \citep{mahadevan2012,mahadevan2014}, we obtain a precise mass measurement of the transiting warm super Neptune, TOI-1728b, orbiting its M0 host star. Observed by the \textit{Transiting Exoplanet Survey Satellite}  \citep[\tess{};][]{Ricker2015}, the inflated planet TOI-1728b is a good candidate for transmission spectroscopy measurements owing to the relatively bright host star (J $\sim 9.6$) and its low planetary density. Our mass measurement precision exceeds the $\sim 20 \%$ (5$\sigma$) recommended by \cite{batalha_precision_2019} for detailed atmospheric characterization. This is important to ensure that the derived atmospheric parameter uncertainties are not dominated by the mass measurement uncertainties.

In Section \ref{sec:obs}, we discuss the observations, which include photometry from \tess{} as well as ground-based photometric follow-up and RV observations with HPF. In Section \ref{sec:stellarparm} we discuss the stellar parameter estimation using HPF spectra and spectral energy distribution (SED) fitting as well as lack of any detectable rotation signal in the photometric data. In Section \ref{sec:jointfit}, we discuss the joint fit of the photometry and the RV observations, followed by a discussion of our upper limit on He 10830 \AA~ absorption in Section \ref{sec:he10830}. In Section \ref{sec:discussion}, we discuss the planetary properties of TOI-1728b with respect to other M dwarf exoplanets. Finally, in Section \ref{sec:summary}, we summarize our results.

\section{Observations}\label{sec:obs}

\subsection{TESS Photometry}\label{seC:tess}
TOI-1728 (TIC 285048486, UCAC4 774-029023, Gaia DR2 1094545653447816064) was observed by \tess{} in Sector 20 from 2019 December 24 to 2020 January 19 at two-minute cadence. It has one transiting planet candidate, TOI-1728.01, with a period of \(\sim3.49\)  days (\autoref{fig:photometry}) that was detected by the \tess{} science processing pipeline \citep{Jenkins2016}. For our subsequent analysis, we used the entire pre-search data conditioned time-series light curves \citep{Ricker2018} available at the Mikulski Archive for Space Telescopes (MAST) for Sector 20. We exclude points marked as anomalous by the \tess{} data quality flags \citep[see Table 28 in][]{Tenenbaum2018}.

\subsection{Ground-based follow-up photometry}
\subsubsection{Perkin 0.43 m} \label{sec:perkins}
We observed a transit of TOI-1728b on the night of 2020 February 22 using the Richard S. Perkin telescope on the campus of Hobart and William Smith Colleges (Geneva, New York, United States).  The 0.43 m (17") f/6.8 Planewave CDK telescope rests on a Paramount equatorial mount with an SBIG 8300 M camera mounted at Cassegrain focus.  The camera detector has an array of $3326 \times 2504$, 5.4 $\micron$ pixels resulting in a $\sim 21 \times 16^\prime$ FOV.    We obtained a series of 92 consecutive images over 5 hours  centered on the target in $1\times1$ binning mode in the Sloan \textit{$r^\prime$} filter.  We defocused moderately (FWHM 3.5\arcsec - 3.8\arcsec) and adopted 180 s exposures which was a compromise that gave both sufficient signal-to-noise on the target and adequate time sampling of the transit.   The guiding was stable and the weather was clear, but our long observing session required us to perform a meridian flip of the mount at 03:23 UT (BJD$_{\rm TDB}$ = 2458902.644523) to observe the egress of the transit.Our observation started at an airmass of 1.14, and after the meridian flip, ended at 1.17.

The data was processed in the standard way with dark subtraction of each image immediately after readout and division by a stacked bias-corrected sky-flat in the \textit{$r^\prime$}-band created from 21 individual sky-flat exposures.   We performed aperture photometry using AstroImageJ \citep{Collins2017} on the processed images.  We tested several different apertures, but ultimately adopted a 15 pixel (5.7\arcsec) radius aperture with a sky annulus of 30-40 pixels in radii (11\arcsec-15\arcsec), which produced the least overall scatter in the final light curve.   The data required only detrending with respect to the meridian flip which occurred at BJD$_{\rm{TDB}}$ = 2458902.644523.  The light curve precision did not improve by detrending with any other parameters, but the position of the star was stable on the detector within $\pm 2$~pixels (0.75\arcsec) on each side of the meridian flip. Since the target TOI-1728 is at a high declination ($\sim 65^{\circ}$), and the observation transited the meridian, the impact of change in airmass was minimal and did not require detrending. The obtained transit is shown in \autoref{fig:photometry}.

\begin{figure*}[!t] 
\centering
\includegraphics[width=0.75\textwidth]{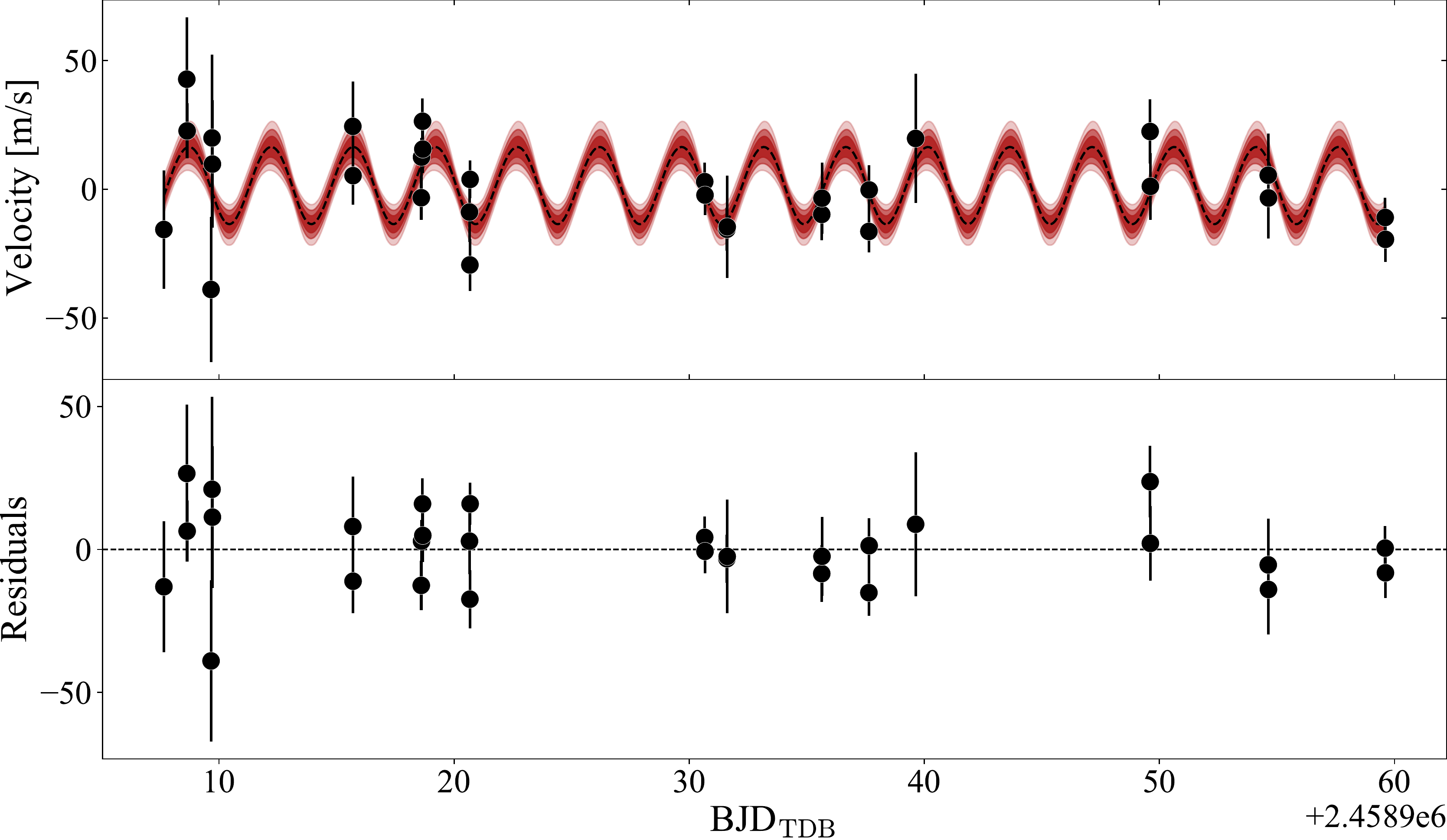}
\caption{Time series of RV observations of TOI-1728 with HPF. The best-fitting model derived from the joint fit to the photometry and RVs is plotted as a dashed line, while the shaded regions denote the \(1\sigma\) (darkest), \(2\sigma\), and \(3\sigma\) range of the derived posterior solution. We also mention the Root Mean Squared Error (RMSE) for the residuals. \explain{Modified plot x axis to JDs.}} \label{fig:rv}
\end{figure*}

\subsubsection{Penn State Davey CDK 0.6 m Telescope} \label{sec:davey}
We observed a transit of TOI-1728b on the night of 2020 February 22 using the 0.6 m telescope located on the roof of the Penn State Davey Laboratory (University Park, Pennsylvania, United States). The telescope was installed in 2014 and has an Apogee/Andor Aspen CG 42 camera, using a CCD42-10 2048 $\times$ 2048 pixel chip from e2v with 13.5 micron pixels. This results in a plate scale of $\sim$ 0.77\arcsec~per pixel and a field-of-view of 24' $\times$ 24'. The observations were made with the Johnson \textit{I} filter and an Engineered Diffuser, which has been described in \cite{Stefansson2017}, with an exposure time of 45 seconds.  Our observations spanned an airmass range of 1.09 to 1.17.

We processed the photometry using AstroImageJ \citep{Collins2017} following the procedures described in \cite{Stefansson2017}. We experimented with a number of different apertures and adopted an object aperture radius of 15 pixels (11.6\arcsec), and inner and outer sky annuli of 25 pixels (19.3\arcsec) and 35 pixels (27.0\arcsec), respectively. These values minimized the standard deviation in the residuals for the data. Following \cite{Stefansson2017}, we added the expected scintillation-noise errors to the photometric error (including photon, readout, dark, sky background, and digitization noise). The transit obtained is shown in \autoref{fig:photometry}.

\subsection{Habitable-zone Planet Finder}\label{sec:hpf}

\begin{deluxetable}{cccc}
\tablecaption{RVs of TOI-1728. All observations have exposure times of 945 s. We include this table in a machine readable format along with the manuscript. \label{tab:rvs}}
\tablehead{\colhead{$\unit{BJD_{TDB}}$}  &  \colhead{RV}   & \colhead{$\sigma$} &\colhead{SNR} \\
           \colhead{}   &  \colhead{$(\unit{m/s})$} & \colhead{$(\unit{m/s})$} }
\startdata
 2458907.65733 &     -15.64 &                22.97 &   58 \\
 2458908.63028 &      42.74 &                24.05 &   51 \\
 2458908.64575 &      22.64 &                10.68 &  122 \\
 2458909.66360 &     -38.93 &                28.24 &   50 \\
 2458909.70634 &      19.92 &                32.39 &   42 \\
 2458909.72115 &       9.77 &                24.77 &   57 \\
 2458915.69333 &      24.41 &                17.40 &   77 \\
 2458915.70361 &       5.27 &                11.23 &  116 \\
 2458918.60459 &      -3.29 &                 8.61 &  146 \\
 2458918.61624 &      12.46 &                 7.46 &  165 \\
 2458918.65883 &      26.40 &                 8.91 &  139 \\
 2458918.67062 &      15.58 &                 9.24 &  139 \\
 2458920.65569 &      -8.86 &                11.60 &  111 \\
 2458920.67280 &     -29.40 &                10.17 &  123 \\
 2458920.68347 &       3.83 &                 7.38 &  170 \\
 2458930.66092 &       2.95 &                 7.34 &  176 \\
 2458930.67270 &      -2.27 &                 7.70 &  165 \\
 2458931.60585 &     -15.54 &                 8.37 &  154 \\
 2458931.61804 &     -14.61 &                19.86 &   69 \\
 2458935.63860 &      -9.78 &                 9.97 &  125 \\
 2458935.64863 &      -3.48 &                13.75 &   90 \\
 2458937.64412 &     -16.40 &                 8.05 &  157 \\
 2458937.65563 &      -0.27 &                 9.62 &  131 \\
 2458939.63169 &      19.73 &                25.13 &   49 \\
 2458949.60579 &      22.42 &                12.48 &  101 \\
 2458949.61659 &       1.12 &                13.00 &   96 \\
 2458954.62964 &       5.46 &                16.19 &   79 \\
 2458954.64087 &      -3.38 &                15.76 &   82 \\
 2458959.60388 &     -10.97 &                 7.67 &  164 \\
 2458959.61532 &     -19.45 &                 8.76 &  143 \\
\enddata
\end{deluxetable}

We observed TOI-1728 using HPF \citep{mahadevan2012,mahadevan2014}, a high-resolution ($R\sim55,000$), NIR (\(8080-12780\)\ \AA) precision RV spectrograph located at the 10 meter Hobby-Eberly Telescope (HET) in Texas. The HET is a fully queue-scheduled telescope with all observations executed in a queue by the HET resident astronomers \citep{shetrone2007}. HPF is a fiber fed instrument with a separate science, sky and simultaneous calibration fiber \citep{kanodia_overview_2018}. It is actively temperature-stabilized and achieves $\sim 1 ~ \rm{mK}$ temperature stability \citep{stefansson2016}. We use the algorithms in the tool \texttt{HxRGproc} for bias noise removal, non-linearity correction, cosmic ray correction, slope/flux and variance image calculation \citep{Ninan2018} of the raw HPF data. We use this variance estimate to calculate the S/N of each HPF exposure (Table \ref{tab:rvs}). Each visit was divided into 2 exposures of 945 seconds each. The median signal-to-noise (S/N) goal was 144 per resolution element. Even though HPF has the capability for simultaneous calibration using a NIR Laser Frequency Comb (LFC), we chose to avoid simultaneous calibration to minimize the impact of scattered calibrator light in the science target spectra. Instead, the stabilized instrument allows us to correct for the well calibrated instrument drift by interpolating the wavelength solution from other LFC exposures from the night of the observations as discussed in \cite{Stefansson2020}. This methodology has been shown to enable precise wavelength calibration and drift correction up to $\sim30$ \cms{} per observation, a value smaller than our estimated per observation RV uncertainty (instrumental + photon noise) for this object of $\sim 10$ \ms{} (in 945 s exposures).

\begin{figure*}[] 
\centering
\includegraphics[width=0.9\textwidth]
{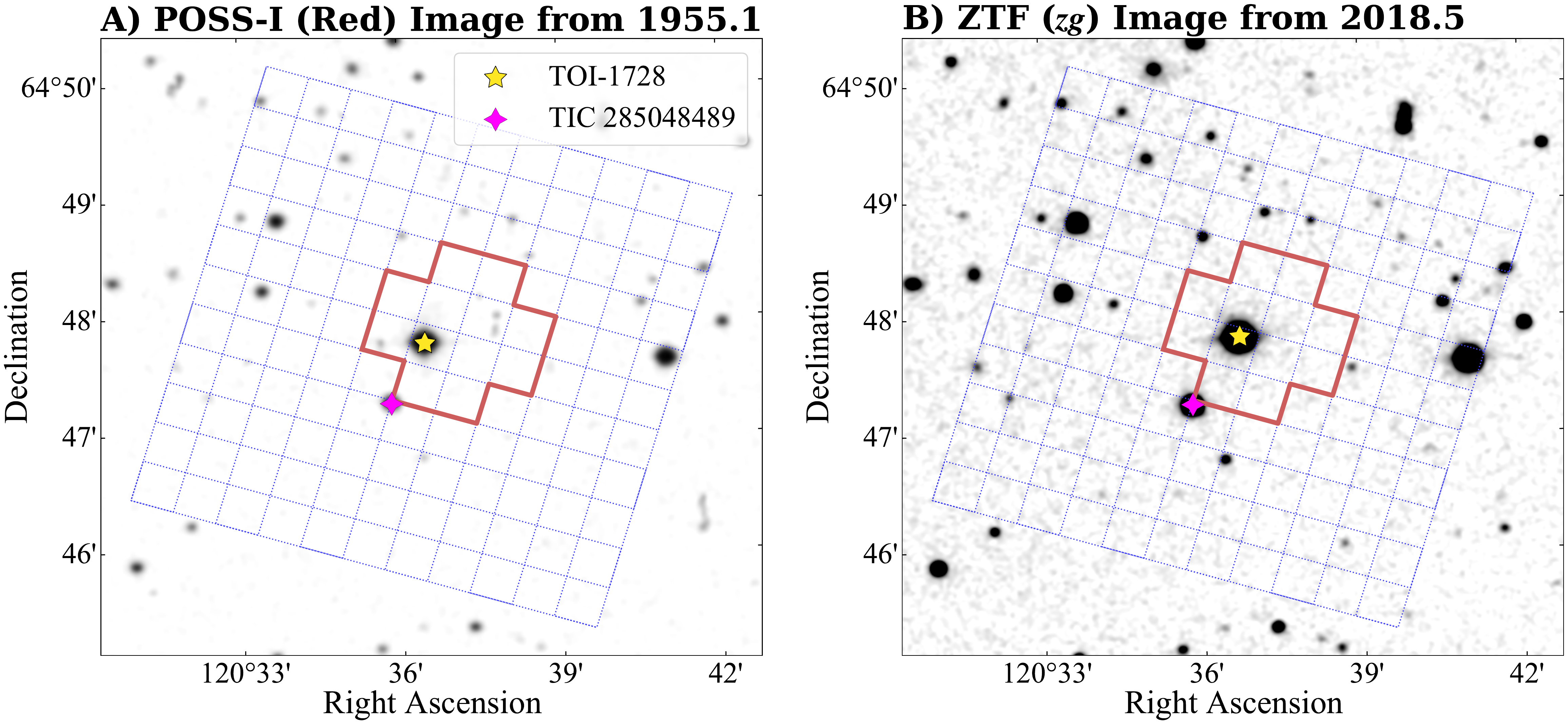}
\caption{\textbf{Panel A} overlays the 11 x 11 pixel \tess{} Sector 20 footprint (blue grid) on a POSS-I red image from 1955. The \tess{} aperture is outlined in red and we highlight our target TOI-1728 and the closest bright star, TIC 285048489. No additional bright stars \(|\Delta G_{RP}|<4\) are contained within the \tess{} aperture. TOI-1728 has moved $\sim 7\arcsec$ between the two epochs. \textbf{Panel B} is similar to Panel A but with a background image from ZTF \(zg\) (4087 \AA -- 5522 \AA) around 2018.5 \citep{Masci2019}. There are no bright targets in the \tess{} aperture that would cause significant dilution to the \tess{} transit.} \label{fig:tess_map}
\end{figure*}

\begin{figure}[!b]
\centering
\includegraphics[width=1.1\columnwidth]
{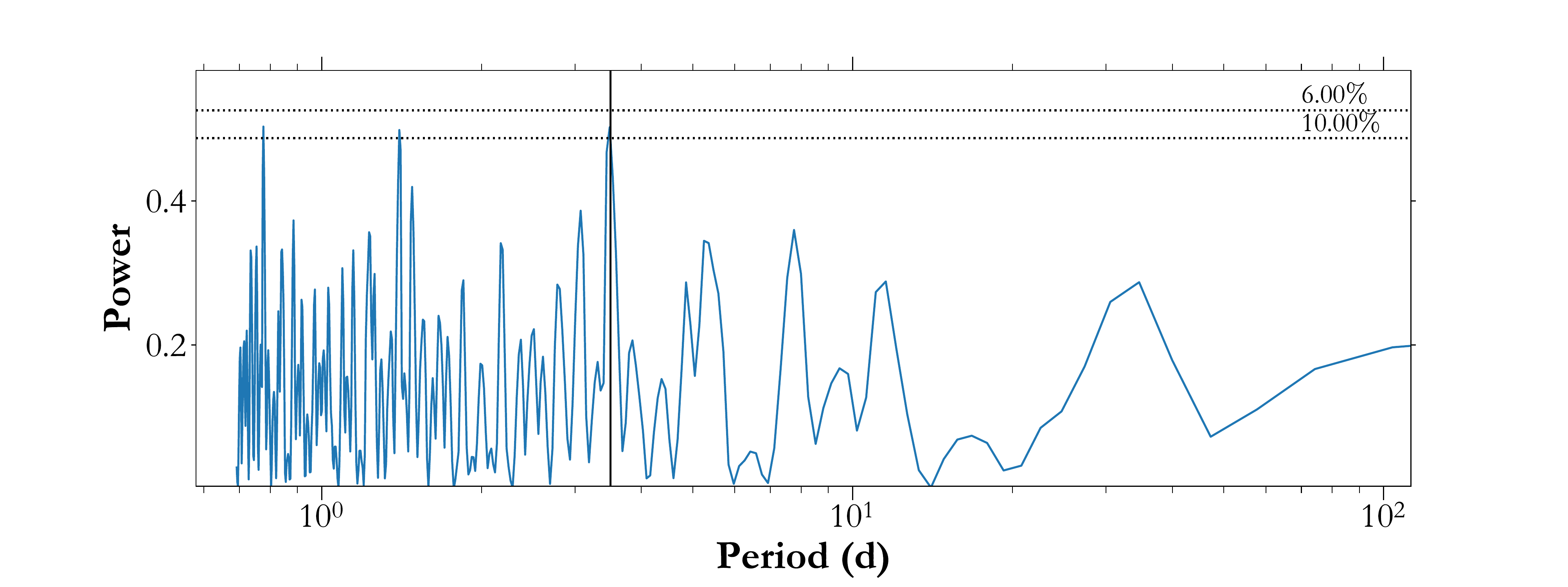}
\caption{\small A GLS periodogram of the HPF RVs showing the $\sim 3.5$ days signal, where the dashed lines mark the False Alarm Probabilities of 6$\%$ and 10$\%$. The vertical line marks the orbital period of TOI-1728b. We do see an additional GLS peak at 1.4 days, however its power is not statistically significant and the phase space for that period is not well sampled by our data.}\label{fig:gls}
\end{figure}

We follow the methodology described in \cite{Stefansson2020} to derive the RVs, by using a modified version of the \texttt{SpEctrum Radial Velocity AnaLyser} pipeline \citep[\texttt{SERVAL};][]{Zechmeister2018}. \texttt{SERVAL} uses the template-matching technique to derive RVs \citep[e.g.,][]{Anglada-Escude2012}, where it creates a master template from the target star observations, and determines the Doppler shift for each individual observation by minimizing the \(\chi^2\) statistic. This master template was generated by using all observed spectra while explicitly masking any telluric regions identified using a synthetic telluric-line mask generated from \texttt{telfit} \citep{Gullikson2014}, a Python wrapper to the Line-by-Line Radiative Transfer Model package \citep{clough2005}. We used \texttt{barycorrpy}, the Python implementation \citep{Kanodia2018} of the algorithms from \cite{wright2014} to perform the barycentric correction. We obtained a total of 36 exposures on this target, of which 3 were excluded from RV analysis since they were taken during a transit of the planet (JD 2458909.61466) and could add a potential systematic due to the Rossiter-McLaughlin effect \citep{1924ApJ....60...15R, 1924ApJ....60...22M, 2018haex.bookE...2T}. Including these RVs in our analysis does not change the results, however, we still choose to exclude these in an abundance of caution. Furthermore, 3 exposures were discarded due to bad weather conditions. The remaining thirty 945 s exposures, are listed in Table \ref{tab:rvs} and plotted in \autoref{fig:rv}.
A generalized Lomb Scargle (GLS) periodogram \citep{2009A&A...496..577Z} on these thirty RV points shows a signal at $\sim 3.5$ days with a False Alarm Probability $\sim 6 \%$ calculated with a bootstrap simulation using \texttt{astropy} which computes periodograms on simulated data given the errors. This period is consistent with the orbital period obtained from the \tess{} photometry.

\section{Ruling out Stellar Companions}
\textbf{Distances 40.0\arcsec to 1.7\arcsec: }
We use light curves derived using the default aperture determined from the \tess{} pipeline. This large aperture typically means that there will be other stars contained within the aperture. Figure \ref{fig:tess_map} presents a comparison of the region contained within the Sector 20 footprint from the Palomar Observatory Sky Survey \citep[POSS-1;][]{Harrington1952,Minkowski1963} image in 1955 and a more recent ZTF \citep{Masci2019} image from 2018. The \tess{} aperture is indicated in red and no additional bright ($|\Delta G_{RP}|<4$) targets are within the aperture. To investigate if any bright background stars are present and diluting the transit, we used \gaia{} DR2 \citep{GaiaCollaboration2018} and searched the $11\times11$ \tess{} pixel grid centered on TOI-1728. \gaia{} detects no bright sources within a radius of 40\arcsec~around our target. Recent results have shown that \gaia{} can recover $> 95 \%$ of  binaries as close as 1.7\arcsec ~for contrasts of up to 4 magnitudes \citep{2018AJ....156..259Z}.  Therefore, we use the lack of a source detection around TOI-1728 to constrain the absence of bright stellar companions at separations from 1.7\arcsec~out to 40.0\arcsec (\(\sim2\) \tess{} pixels). 

\textbf{Distances $>$ 40.0\arcsec: }
The closest neighbor detected in \gaia{} DR2, TIC 285048489 (\gaia{} DR2 1094545619088078208; $T=14.12$, $G_{RP}=14.06$, $\Delta G_{RP} = 3.26$\footnote{We use the \gaia{} DR2 $G_{RP}$ bandpass ($\sim$ 7700 \AA  ~ to 10600 \AA  ~) magnitudes as a proxy for the \tess{} bandpass ($\sim$ 5800\AA  ~ to 11100 \AA) magnitudes.}), is contained outside the aperture. Both TOI-1728 and TIC 285048489 have small proper motions and only our target was contained in the aperture when \tess{} observed this region. The centroid for TIC 285048489 lies just outside aperture such that we can expect a small amount of dilution in the \tess{} light curve due to the presence of TIC 285048489, which we include as a dilution term in our joint analysis of the photometry and velocimetry (Section \ref{sec:jointfit}). This star is at a sky projected separation of \(41\arcsec\) from TOI-1728 and is excluded in apertures used for the ground-based photometry described in Sections \ref{sec:perkins} and \ref{sec:davey}.

\textbf{Distances $<$ 1.7\arcsec: } 
In a span of $\sim 60$ years, between the recent ZTF image (2018.5) and the POSS-1 image (1955.1),  TOI-1728 has had a sky projected motion of $\sim$ 7\arcsec. The POSS-1 plate images were taken with Eastman 103a-E spectroscopic plates in conjunction with a No. 160 red plexiglass filter with a bandpass between 6000 -- 6700 \AA, and have a limiting magnitude of $R \sim 19$ \citep{Harrington1952}. Based on the POSS-1 field of view of this region (\autoref{fig:tess_map}), we rule out background objects at the present position for TOI-1728 with contrast $\Delta R < 6$ \citep[TOI-1728 Johnson $R = 11.9$, ][]{2012yCat.1322....0Z}. 

\begin{figure}[!t]
\centering
\includegraphics[width=0.9\columnwidth]
{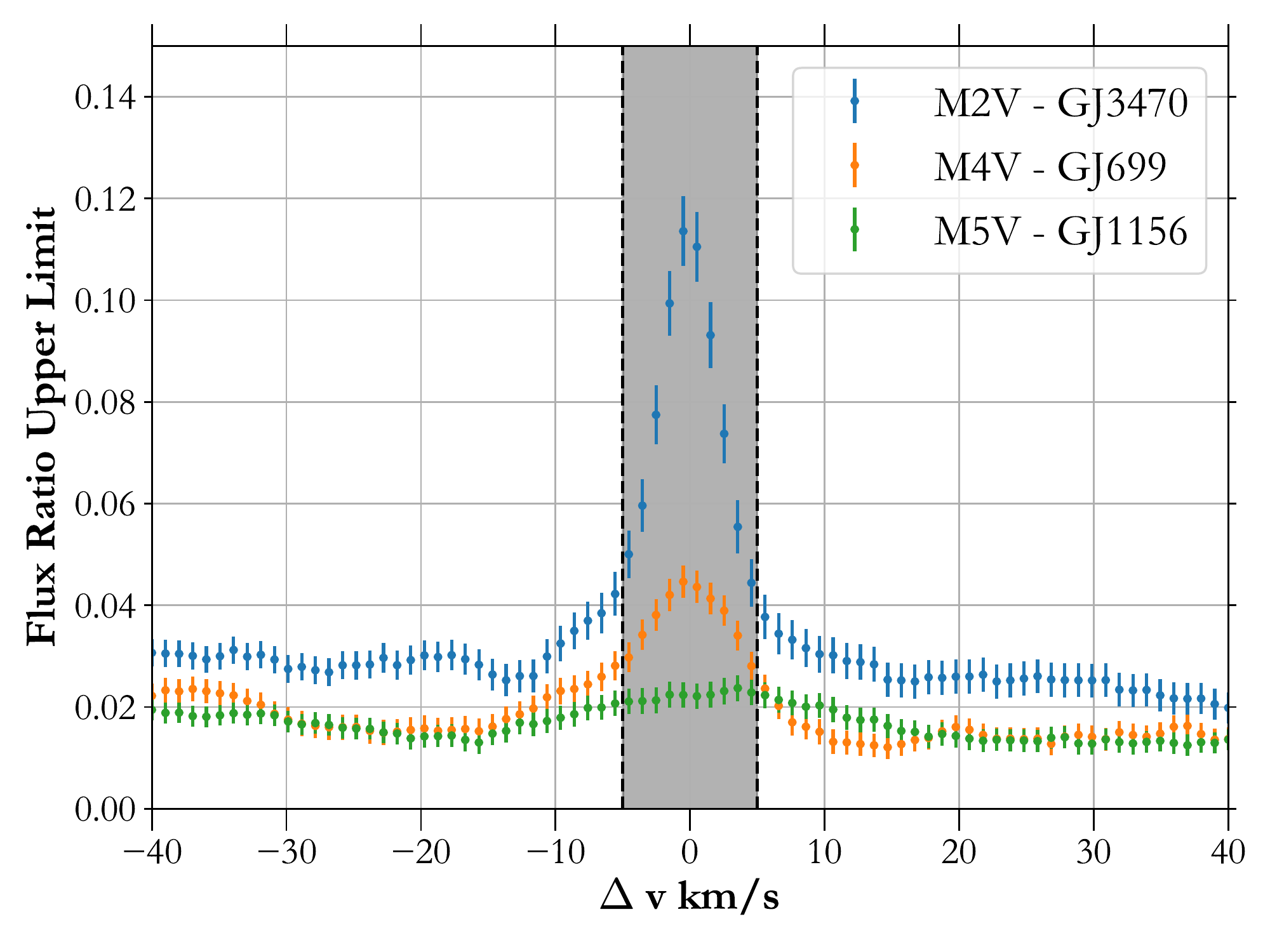}
\caption{\small \added{Flux upper} limits placed on the flux ratio of a secondary companion to an M0 template (GJ488) as a function of $\Delta \rm{v}$, obtained using HPF order 69 spanning $\sim 8780 - 8890$ \AA. We include the 1 $\sigma$ erorr bars, and shade the region corresponding to $\pm 5 $ \kms{}. Using this, we place a conservative upper limit of an unresolved secondary of flux ratio = 0.05 for $|\Delta v| > 5$~ \kms{}. \explain{Updated y label}}\label{fig:signalinjection}
\end{figure}

\begin{figure*}[!t]  
\centering
\includegraphics[width=0.9\textwidth]{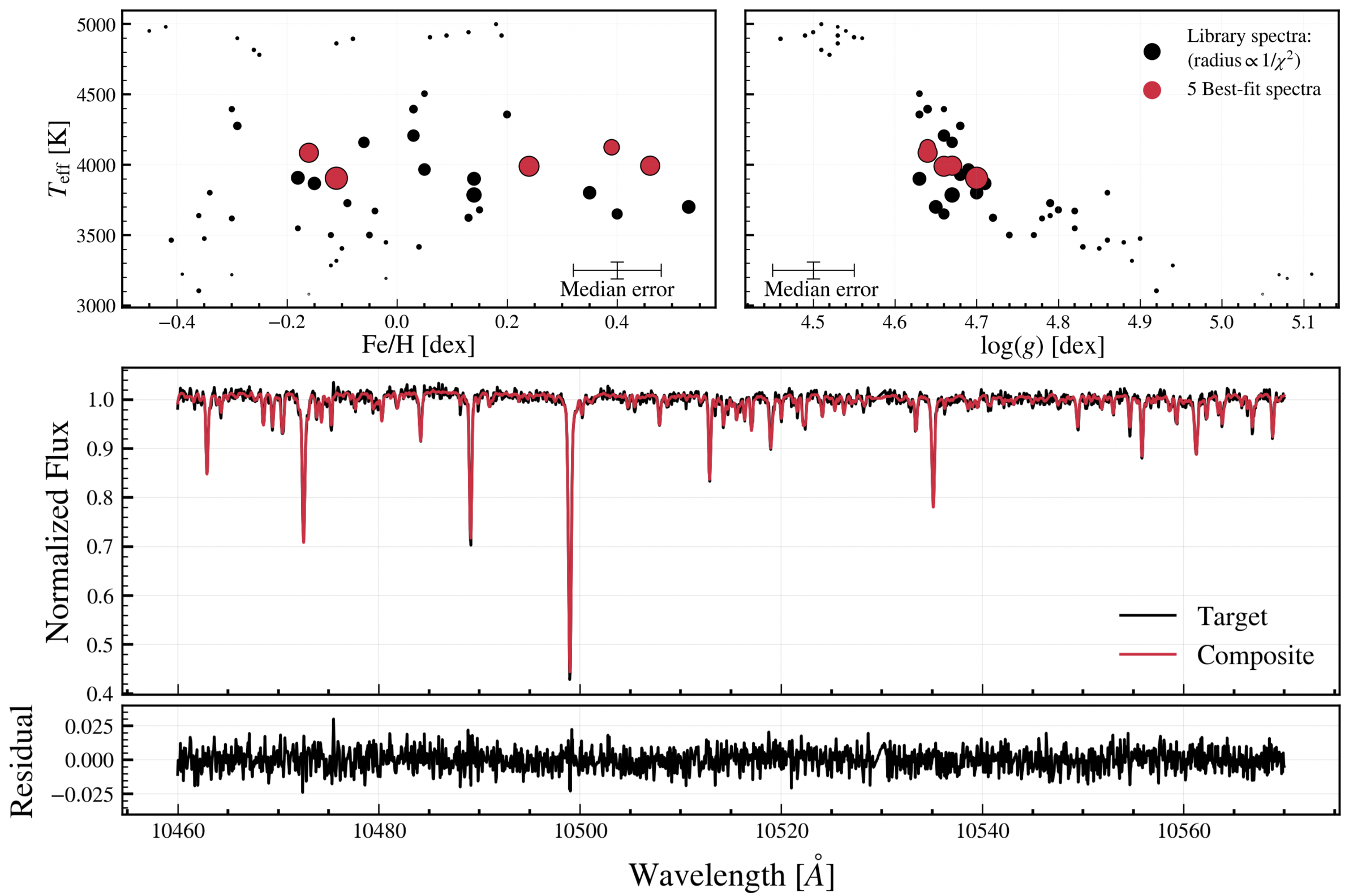}
\caption{\textbf{Top panels:} Best-fit library stars to TOI-1728 showing the \teff{} of the library stars as a function of [Fe/H] (left) and \(\log g\) (right). The radius of each data point is inversely proportional to the calculated $\chi^2$ initial value where we compare the target (TOI-1728) to the reference star spectra, so larger points show a lower $\chi^2$ initial value and thus indicate a better fit. Highlighted in the red dots are the 5 best matching stars which we use to construct a linear combination composite spectrum to derive our final stellar parameters.  Target spectrum (black) compared to our best-fit linear combination composite spectrum (red). \textbf{Bottom:} Residuals from the fit are shown.} 
\label{fig:specmatch}
\end{figure*}

\textbf{Unresolved Bound companions: }
The host stellar density is constrained from the transit fit \citep{seager_unique_2003} to be consistent with that obtained from the SED fit for an M0 host star (Section \ref{sec:stellarparams}). Furthermore, we place limits on a spatially unresolved bound companion by quantifying the lack of flux from a secondary stellar object in the HPF spectra.  We parameterize the TOI-1728 spectra as a linear combination of a primary of M0 spectral type (GJ 488\footnote{GJ488 was chosen for the M0 template since it represented one of the closest match to the TOI-1728 spectra (\autoref{fig:specmatch}).}), and a secondary stellar companion, where the flux ratio is given by the ratio of these coefficients (\autoref{eq:secondary}).

\begin{equation} \label{eq:secondary}
\begin{split}
S_{\rm{TOI-1728}} ~ &= ~ (1-x)~S_{\rm{M0}} + (\textit{x})S_{\rm{S}} \\
F &= \frac{x}{1-x}
\end{split}
\end{equation}

\noindent where $S_{\rm{TOI-1728}}$ is the TOI-1728 spectra,  $S_{\rm{M0}}$ is the primary spectra of M0 spectral type (GJ488), and $S_{\rm{S}}$ represents the secondary spectra (bound companion). Here \textit{x} is the parameter we fit, and \textit{F} is the flux ratio of the secondary to the primary which is plotted in \autoref{fig:signalinjection}. We shift the secondary spectra in velocity space and then add that to the primary M0 template in order to match the TOI-1728 spectra. We limit this fitting to secondary companions of stellar types fainter than M0, i.e. ($>$ M0), and obtain flux limits for secondary stellar companions of M2 (GJ3470),  M4 (GJ699) and M5 (GJ1156) spectral types. Bound stellar secondary companions of spectral type later than M5 would be too faint to cause appreciable dilution in the transit light curve, and their impact would be below our current precision on the radius estimate. Our RV residuals with a baseline of $\sim$ 50 days do not show a significant trend (linear or otherwise), and therefore we estimate the impact of an unresolved bound companion on our estimated planetary parameters should be negligible. We use high S/N spectra observed by HPF of these stars. We do not perform this procedure in the entire spectrum, but in individual orders due to some of the orders being dominated by telluric absorption. The flux ratio estimates are consistent across the non telluric dominated orders, and for conciseness we present here the results from HPF order 69 spanning $\sim 8780 - 8890$ \AA. We place a conservative upper limit for a secondary of flux ratio = 0.05  or $\Delta \rm{mag} \simeq 3.25$ at velocity offsets ($|\Delta v|$)  $> 5$ \kms{}. As shown in \autoref{fig:signalinjection}, we do not see significant flux contamination at ($|\Delta v|$)  $> 5$ \kms{}.  We perform this fitting for the flux ratio for velocity offsets from 5 to 40 \kms{}, where the lower limit is approximately HPF's spectral resolution ($\sim 5.5 \rm{km/s}$). At velocity offsets $<$ 5 \kms{}, the degeneracy between the primary and secondary spectra makes it difficult to place meaningful flux ratio constraints.

\section{Stellar Parameters} \label{sec:stellarparm}

\subsection{Spectroscopic Parameter Estimation}
To measure spectroscopic $T_{\mathrm{eff}}$, $\log g$, and [Fe/H] values of the host star, we use the HPF spectral matching methodology from \cite{Stefansson2020}, which is based on the \texttt{SpecMatch-Emp} algorithm from \cite{yee_precision_2017}. In short, we compare our high resolution HPF spectra of TOI-1728 to a library of high S/N as-observed HPF spectra. The library consists of slowly-rotating reference stars with well characterized stellar parameters from \cite{yee_precision_2017}. 

To perform the comparison, we shift the observed target spectrum to a library wavelength scale and rank all of the targets in the library using a $\chi^2$ goodness-of-fit metric. Following the initial $\chi^2$ minimization step, we pick the five best matching reference spectra (in this case: GJ 1172, GJ 488, BD+29 2279, HD 88230, and HD 28343) to construct a linear combination weighted composite spectrum to better match to the target spectrum (see Figure \ref{fig:specmatch}). In this step, each of the five stars receives a best-fit weight coefficient. We then assign the target stellar parameter $T_{\mathrm{eff}}$, $\log g$, and Fe/H values as the weighted average of the five best stars using the best-fit weight coefficients. Our final parameters are listed in \autoref{tab:stellarparam}, using the cross-validation error estimates from \cite{Stefansson2020}. As an additional check, we performed the library comparison using 6 other HPF orders which have low levels of tellurics, all of which resulted in consistent stellar parameters. Lastly, during both optimization steps, we note that we account for any potential $v \sin i$ broadening by artificially broadening the library spectra with a $v \sin i$ broadening kernel \citep{gray1992} to match the rotational broadening of the target star. For TOI-1728 no significant rotational broadening was needed, and we thus place an upper limit of $v \sin i < 2 \unit{km/s}$, which is the lower limit of measurable $v \sin i$ values given HPF's spectral resolving power of $R \sim 55,000$.


\begin{figure}[!h] 
\centering
\includegraphics[width=\columnwidth]
{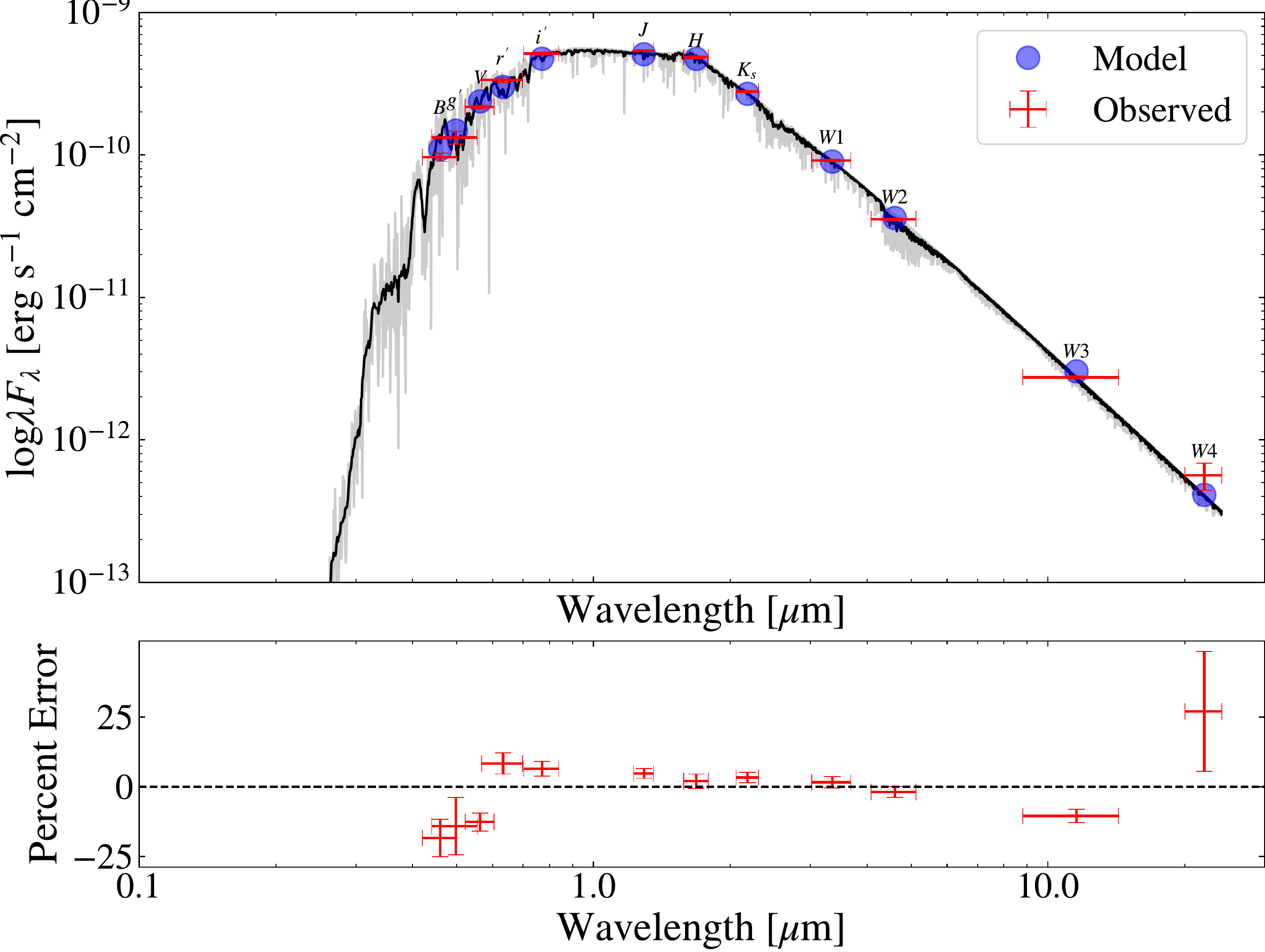}
\caption{The SED of TOI-1728. The grey line is the raw BT-NextGen model and black line is the
model smoothed with a boxcar average of 10 points. The SED was fit with \texttt{EXOFASTv2} using the distance inferred from \gaia{} DR2. The error bars in wavelength reflect the bandwidth of the respective
photometric filter and the error bars in flux reflect the measurement uncertainty. The blue circles are the points on the best-fitting model corresponding to the midpoint of each photometric filter. The
bottom panel shows the percent error between the best-fitting model and the observed magnitudes. The resulting stellar parameters are summarized in \autoref{tab:stellarparam}. \explain{Updated Figure}} \label{fig:sed}
\end{figure}

\subsection{Model-Dependent Stellar Parameters}\label{sec:stellarparams}

In addition to the spectroscopic stellar parameters derived above, we use \texttt{EXOFASTv2} \citep{Eastman2019} to model the SED of TOI-1728 (\autoref{fig:sed}) to derive model-dependent constraints on the stellar mass, radius, and age of the star. For the SED fit, \texttt{EXOFASTV2} uses the the BT-NextGen stellar atmospheric models \citep{Allard2012}.  We assume Gaussian priors on the (i) 2MASS \(JHK\) magnitudes, (ii) SDSS \(g^\prime r^\prime i^\prime\) and Johnson \(B\) and \(V\) magnitudes from APASS, (iii) \textit{Wide-field Infrared Survey Explorer} magnitudes $W1$, $W2$, $W3$, and $W4$, \citep[][]{Wright2010}, (iv) spectroscopically-derived host star effective temperature, surface gravity, and metallicity, and (v) distance estimate from \cite{Bailer-Jones2018}. We apply a uniform prior on the visual extinction and place an upper limit using estimates of Galactic dust by \cite{Green2019} (Bayestar19) calculated at the distance determined by \cite{Bailer-Jones2018}. We convert the Bayestar19 upper limit to a visual magnitude extinction using the \(R_{v}=3.1\) reddening law from \cite{Fitzpatrick1999}. We use \texttt{GALPY} \citep{2015ApJS..216...29B} to calculate the UVW velocities, which along with the BANYAN tool \citep{2018ApJ...856...23G} classify TOI-1728 as a field star in the thin disk \citep{2014A&A...562A..71B}.

The stellar priors and derived stellar parameters with their uncertainties are listed in Table \ref{tab:stellarparam}. We use the \teff{} estimate of $= 3980^{+31}_{-32}$ K to classify TOI-1728 as an M star (\teff{} $< 4000 K$), which is consistent with results obtained from the PPMXL catalogue \citep[][]{2010AJ....139.2440R,  2013MNRAS.435.2161F}.

\begin{deluxetable*}{lccc}
\tablecaption{Summary of stellar parameters for TOI-1728. \label{tab:stellarparam}}
\tablehead{\colhead{~~~Parameter}&  \colhead{Description}&
\colhead{Value}&
\colhead{Reference}}
\startdata
\multicolumn{4}{l}{\hspace{-0.2cm} Main identifiers:}  \\
~~~TOI & \tess{} Object of Interest & 1728 & \tess{} mission \\
~~~TIC & \tess{} Input Catalogue  & 285048486 & Stassun \\
~~~2MASS & \(\cdots\) & J08022653+6447489 & 2MASS  \\
~~~Gaia DR2 & \(\cdots\) & 1094545653447816064 & Gaia DR2\\
\multicolumn{4}{l}{\hspace{-0.2cm} Equatorial Coordinates, Proper Motion and Spectral Type:} \\
~~~$\alpha_{\mathrm{J2000}}$ &  Right Ascension (RA) & 08:02:26.55 & Gaia DR2\\
~~~$\delta_{\mathrm{J2000}}$ &  Declination (Dec) & +64:47:48.93 & Gaia DR2\\
~~~$\mu_{\alpha}$ &  Proper motion (RA, \unit{mas/yr}) & $104.078\pm0.044$ & Gaia DR2\\
~~~$\mu_{\delta}$ &  Proper motion (Dec, \unit{mas/yr}) & $52.919\pm0.046$ & Gaia DR2 \\
~~~$d$ &  Distance in pc  & $60.80^{+0.14}_{-0.13}$ & Bailer-Jones \\
~~~\(A_{V,max}\) & Maximum visual extinction & 0.01 & Green\\
\multicolumn{4}{l}{\hspace{-0.2cm} Optical and near-infrared magnitudes:}  \\
~~~$B$ & Johnson B mag & $13.693\pm0.073$ & APASS\\
~~~$V$ & Johnson V mag & $12.400\pm0.035$ & APASS\\
~~~$g^{\prime}$ &  Sloan $g^{\prime}$ mag  & $13.086\pm0.112$ & APASS\\
~~~$r^{\prime}$ &  Sloan $r^{\prime}$ mag  & $11.788\pm0.041$ & APASS \\
~~~$i^{\prime}$ &  Sloan $i^{\prime}$ mag  & $11.096\pm0.029$ & APASS \\
~~~$J$ & $J$ mag & $9.642\pm0.019$ & 2MASS\\
~~~$H$ & $H$ mag & $8.953\pm0.028$ & 2MASS\\
~~~$K_s$ & $K_s$ mag & $8.803\pm0.020$ & 2MASS\\
~~~$W1$ &  WISE1 mag & $8.691\pm0.022$ & WISE\\
~~~$W2$ &  WISE2 mag & $8.742\pm0.021$ & WISE\\
~~~$W3$ &  WISE3 mag & $8.598\pm0.026$ & WISE\\
~~~$W4$ &  WISE4 mag & $8.250\pm0.234$ & WISE\\
\multicolumn{4}{l}{\hspace{-0.2cm} Spectroscopic Parameters$^a$:}\\
~~~$T_{\mathrm{eff}}$ &  Effective temperature in \unit{K} & $3975\pm77$& This work\\
~~~$\mathrm{[Fe/H]}$ &  Metallicity in dex & $0.09\pm0.13$ & This work\\
~~~$\log(g)$ & Surface gravity in cgs units & $4.67\pm0.05$ & This work\\
\multicolumn{4}{l}{\hspace{-0.2cm} Model-Dependent Stellar SED and Isochrone fit Parameters$^b$:}\\
~~~$T_{\mathrm{eff}}$ &  Effective temperature in \unit{K} & $3980^{+31}_{-32}$ & This work\\
~~~$\mathrm{[Fe/H]}$ & Metallicity in dex & $0.25\pm0.10$ & This work \\
~~~$\log(g)$ &  Surface gravity in cgs units & $4.657\pm{0.017}$ & This work \\
~~~$M_*$ &  Mass in $M_{\odot}$ & $0.646_{-0.022}^{+0.023}$ & This work\\
~~~$R_*$ &  Radius in $R_{\odot}$ & $0.6243_{-0.0097}^{+0.010}$ & This work\\
~~~$L_*$ &  Luminosity in $L_{\odot}$ & $0.088\pm{0.002}$ & This work\\
~~~$\rho_*$ &  Density in $\unit{g/cm^{3}}$ & $3.78\pm{0.19}$ & This work\\
~~~Age & Age in Gyrs & $7.1\pm4.6$ & This work\\
~~~$A_v$ & Visual extinction in mag & $0.0050_{-0.0035}^{+0.0034}$ & This work\\
\multicolumn{4}{l}{\hspace{-0.2cm} Other Stellar Parameters:}           \\
~~~$v \sin i_*$ &  Rotational velocity in \unit{km/s}  & $<2$ & This work\\
~~~$\Delta RV$ &  ``Absolute'' radial velocity in \unit{km/s} & $-43.2 \pm 0.3$ & Gaia DR2\\
~~~$U, V, W$ &  Galactic velocities in \unit{km/s} &  $53.8 \pm 0.2, -9.8 \pm 0.1, 1.8 \pm 0.15$ & This work\\
\enddata
\tablenotetext{}{References are: Stassun \citep{Stassun2018}, 2MASS \citep{2003yCat.2246....0C}, Gaia DR2 \citep{GaiaCollaboration2018}, Bailer-Jones \citep{Bailer-Jones2018}, Green \citep{Green2019}, APASS \citep{Henden2018}, WISE \citep{Wright2010}}
\tablenotetext{a}{Derived using the HPF spectral matching algorithm from \cite{Stefansson2020}.}
\tablenotetext{b}{{\tt EXOFASTv2} derived values using MIST isochrones with the \gaia{} parallax and spectroscopic parameters in $a$) as priors.}
\end{deluxetable*}

\subsection{No Detectable Stellar rotation signal}
To estimate the stellar rotation period, we accessed the publicly available data from the Zwicky Transient Facility \citep[ZTF,][]{Masci2019} and ASAS-SN \citep{Kochanek2017} for this target to perform a Lomb-Scargle \citep{1976Ap&SS..39..447L, 1982ApJ...263..835S} periodogram analysis. 
We do not detect any rotation signal present in the photometry. The photometry spans JD 2458202 -- 2458868 ($\sim$ 660 days) for ZTF (\(zg\)) and JD 2455951 -- 2458934 ($\sim$ 3000 days) for ASAS-SN (in the \textit{V} and \textit{g} band). We repeated this on the photometry from the \tess{} PDCSAP pipeline \citep{2012PASP..124.1000S, Ricker2018} and do not see any statistically significant signals which would suggest potential rotation modulation signal in the photometry. 

The lack of photometric rotational modulation in the long baseline photometry suggests an inactive star, and this claim is further bolstered by the lack of emission or any detectable temporal changes in the cores of the Calcium II NIR triplet \citep{mallik_ca_1997, cincunegui_h_2007, martin_ca_2017} in the HPF spectra. This combination of low stellar activity and $v \sin i$ upper limit of 2 \kms{} (Section \ref{sec:stellarparm}) suggest a slow rotating old and inactive star, which is consistent with our age estimate from the \texttt{EXOFASTv2} fit of 7 $\pm$ 4.6 Gyr.

\section{Data Analysis}

\subsection{Joint fitting of photometry and RVs}\label{sec:jointfit}

We conduct a joint data analysis of all photometry (\tess{} and ground-based), and the RVs using two independent tools: \texttt{juliet} \citep{Espinoza2019} and \texttt{exoplanet} \citep{exoplanet:exoplanet}. The \texttt{juliet} package uses the dynamic nest-sampling algorithm \texttt{dynesty} \citep{Speagle2019} for parameter estimation, whereas \texttt{exoplanet} uses the \texttt{PyMC3} Hamiltonian Monte Carlo package \citep{exoplanet:pymc3}. We confirm that the results of the two independent methods are consistent to 1-$\sigma$ for all derived planetary properties, and present the results of the \texttt{juliet} analysis in this paper for brevity.

\texttt{juliet} uses \texttt {batman} \citep{Kreidberg2015} to model the photometry and \texttt{radvel} \citep{Fulton2018} to model the velocimetry. The RV model is a standard Keplerian model while the photometric model is based on the analytical formalism of \cite{Mandel2002} for a planetary transit and assumes a quadratic limb-darkening law. In the photometric model we include a dilution factor, \(D\), to represent the ratio of the out-of-transit flux of TOI-1728 to that of all the stars within the \tess{} aperture. We assume that the ground based photometry has no dilution, since we use the ground based transits to estimate the dilution in the \tess{} photometry. We assume the transit depth is identical in all bandpasses and use our ground-based transits to determine the dilution required in the \tess{} data to be $D_{TESS}$ $= 0.860\pm0.045$; including which increases the radius from $4.71_{-0.11}^{+0.14} R_{\oplus}$ to $5.04_{-0.17}^{+0.16} R_{\oplus}$.  We also set a prior on the stellar density using the value determined from our \texttt{EXOFASTv2} SED fit. For both the photometry and RV modeling, we include a simple white-noise model in the form of a jitter term that is added in quadrature to the error bars of each data set.

The photometric model includes a Gaussian Process (GP) model to account for any correlated noise behavior in the \tess{} photometry. We use the \texttt{celerite} implementation of the quasi-periodic covariance function \citep[Equation 56 in][]{Foreman-Mackey2017} available in \texttt{juliet}. The \texttt{celerite} covariance parameters are chosen such that the combination of parameters replicates the same behavior as a quasi-periodic kernel \citep{Rasmussen2006}. We follow the example in \cite{Foreman-Mackey2017} and set broad uniform priors on the hyperparameters for the GP.


\texttt{juliet} uses the default stopping criterion from \texttt{dynesty} \citep{Speagle2019}, which stops the run when the estimated contribution of the remaining prior volume to that of the total evidence falls below a preset threshold. For this analysis, we defined that threshold to be 1\%, i.e. the run stopped when the remaining prior volume was lower than 1\% of the total evidence. Furthermore, we verify this result using an Hamiltonian Monte Carlo parameter estimation implemented in \texttt{PyMC3} \citep{exoplanet:pymc3} under \texttt{exoplanet} \citep{exoplanet:exoplanet}, which uses the Gelman-Rubin statistic \citep[$\hat{\text{R}} \le 1.1$;][]{Ford2006} to check for convergence.

\begin{figure}[!t] 
\centering
\includegraphics[width=0.45\textwidth]{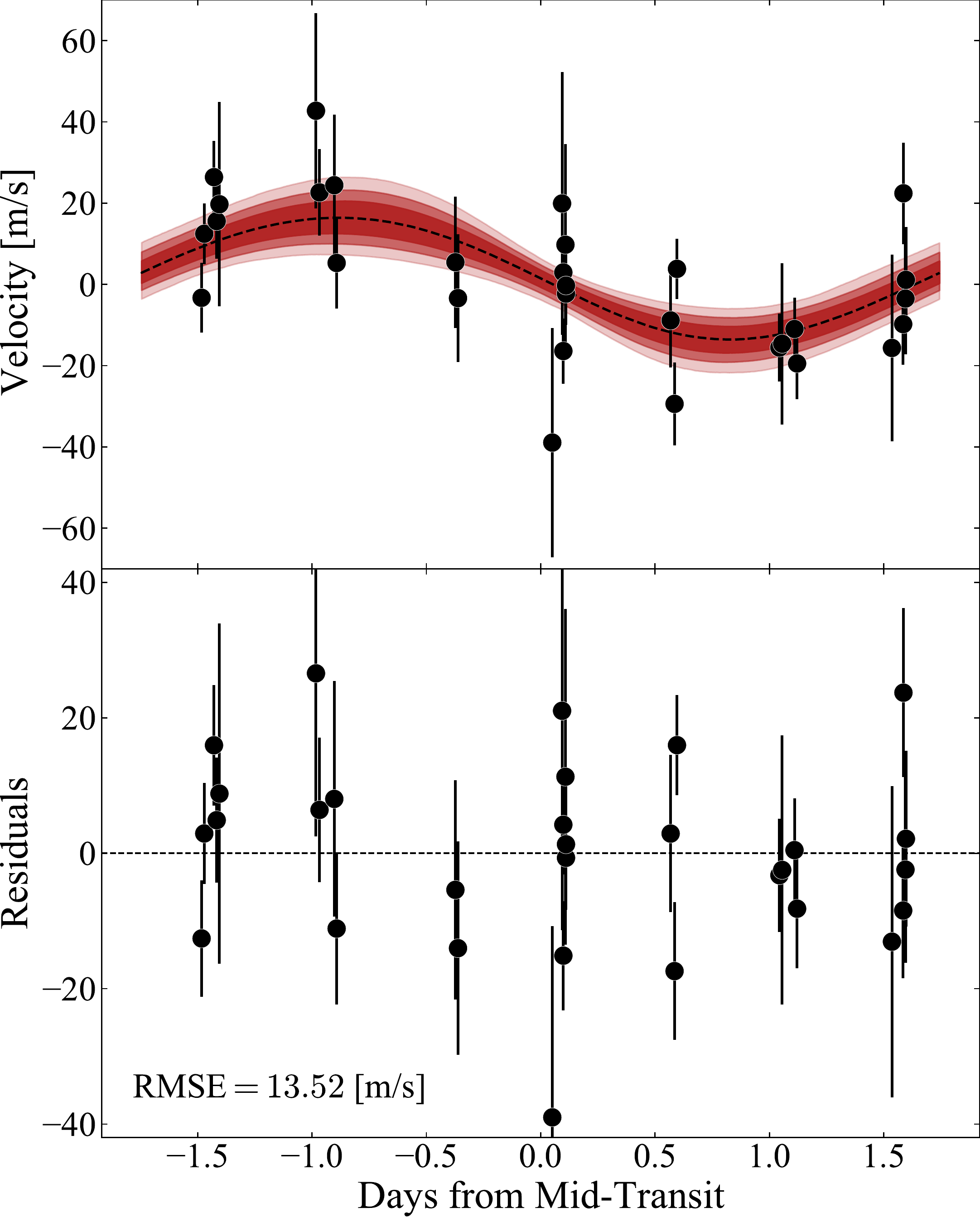}
\caption{The RV observations phase folded on the best-fit orbital period obtained from the joint fit from Section \ref{sec:jointfit}. \added{The best-fitting model is plotted as a dashed line while the shaded regions denote the \(1\sigma\) (darkest), \(2\sigma\), and \(3\sigma\) range of the derived posterior solution.}} \label{fig:phasedrvs}
\end{figure}

\begin{figure*}[!h] 
\centering
\includegraphics[width=1\textwidth]
{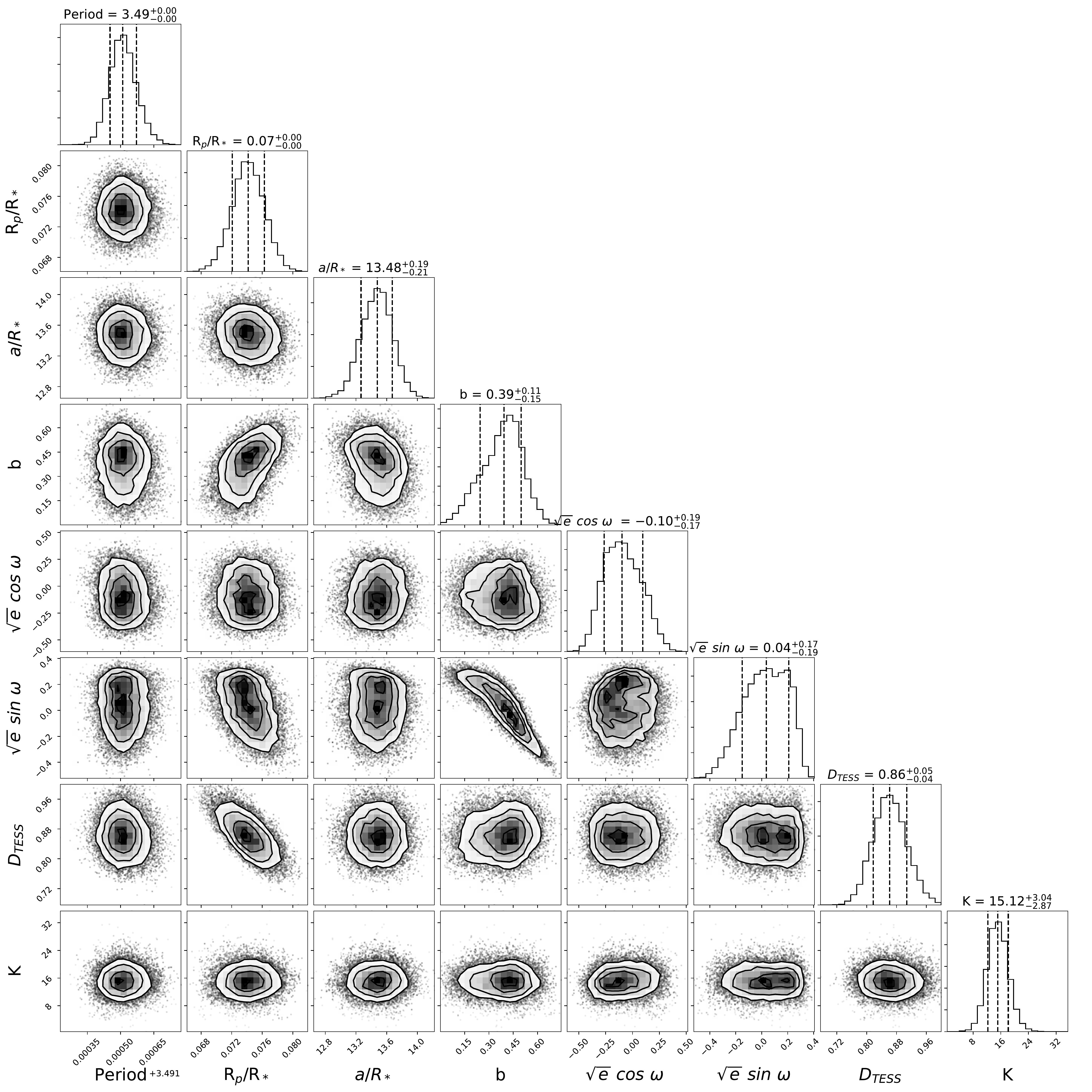}
\caption{Corner plot generated using the posteriors from dynamic nested sampling from \texttt{juliet}. There are significant correlations between $b$ and  $\sqrt{e}~$sin$ ~\omega$, and between $b$,  $R_{p}/R_{*}$, and $D_{TESS}$.  These cause the marginal posteriors for b and $\sqrt{e}~$sin$ ~\omega$ to be significantly skewed. We do not find scientifically significant correlations between the  remaining parameters. We include the posteriors as a data file along with the manuscript. Plot generated using \texttt{corner.py} \citep{corner}.} \label{fig:corner}
\end{figure*}

In \autoref{fig:photometry} we show the photometry and the best fit transit model, in \autoref{fig:rv} we show the RV time series and the model from this joint fit, showing the residuals in the bottom panel \autoref{fig:phasedrvs} shows the phase folded RVs.  Table \ref{tab:priors1} lists the priors used in \texttt{juliet},  Table \ref{tab:planetprop} provides a summary of the inferred system parameters and respective confidence intervals, and \autoref{fig:corner} shows a corner plot of the posteriors. The data reveal a companion having a radius of $5.05_{-0.17}^{+0.16}$ R$_{\oplus}$ and mass $26.78_{-5.13}^{+5.43}$ M$_{\oplus}$.

\begin{deluxetable}{llc}
\tablecaption{Summary of priors used for the three joint transit and RV fits performed. $\mathcal{N}(\mu,\sigma)$ denotes a normal prior with mean $\mu$, and standard deviation $\sigma$; $\mathcal{U}(a,b)$ denotes a uniform prior with a start value $a$ and end value $b$, $\mathcal{J}(a,b)$ denotes a Jeffreys prior  truncated between a start value $a$ and end value $b$. A Gaussian prior on the stellar density was placed for all fits. The dilution parameters in \texttt{juliet} were fixed to 1 for all ground based transit observations.\label{tab:priors1}}
\tablehead{\colhead{Parameter}&  \colhead{Description}                             & \colhead{Model}} 
\startdata	
\hline
\multicolumn{3}{l}{\hspace{-0.3cm} Orbital Parameters:}           \\                    
$P$ (days)                    &  Orbital Period                                    & $\mathcal{N}(3.48,0.1)$        \\ 
$T_C$                         &  Transit Midpoint $(\mathrm{BJD_{TDB}})$ & $\mathcal{N}(2458843.276,0.1)$   \\ 
$R_{p}/R_{*}$                 &  Scaled Radius               		     & $\mathcal{U}(0,1)$                     \\
$a/R_*$                       &  Scaled Semi-major axis                  & $\mathcal{J}(1,90)$                    \\ 
$b$                           &  Impact Parameter                        & $\mathcal{U}(0,1)$                     \\
$\sqrt{e}~$cos$ ~\omega$                             &  e - $\omega$ parameterization     &  $\mathcal{U}(-1,1)$ \\ 
$\sqrt{e}~$sin$ ~\omega$                             &  e - $\omega$ parameterization     &  $\mathcal{U}(-1,1)$ \\ 
$K$                           &  RV semi-amplitude ($\unit{m/s}$)                  & $\mathcal{J}(0.001,100)$        \\ 
\multicolumn{3}{l}{\hspace{-0.3cm} Other constraints:}           \\
$\rho_*$                      &  Stellar density ($\unit{g\:cm^{-3}}$)             & $\mathcal{N}(3.78,0.19)$       \\ 
\multicolumn{3}{l}{\hspace{-0.3cm} Jitter and other instrumental terms:}           \\
$u_1^{a}$                   &  Limb-darkening parameter                          & $\mathcal{U}(0,1)$                     \\ 
$u_2^{a}$                   &  Limb-darkening parameter                          & $\mathcal{U}(0,1)$                     \\ 
$\sigma_{\mathrm{phot}}$$^b$  &  Photometric jitter ($\unit{ppm}$)                 & $\mathcal{J}(10^{-6},~5000)$      \\ 
$\mu_{\mathrm{phot}}$$^b$     &  Photometric baseline                              & $\mathcal{N}(0,0.1)$                   \\ 
$D_{TESS}$     &  Photometric Dilution                              & $\mathcal{U}(0,1)$                   \\ 
$D_\textrm{Davey}$     &  Photometric Dilution                              & Fixed(1)                  \\ 
$D_\textrm{Perkin}$     &  Photometric Dilution                              & Fixed(1)                  \\ 
$\sigma_{\mathrm{HPF}}$       &  HPF RV jitter ($\unit{m/s}$)                      & $\mathcal{J}(0.001,1000)$            \\ 
$\gamma$          &  Systemic velocity ($\unit{m/s}$)                      & $\mathcal{N}(0,100)$         \\ 
$dv/dt$  &  HPF RV trend ($\unit{mm/s/day}$)   & $\mathcal{N}(0,500)$         \\ 
\multicolumn{3}{l}{\hspace{-0.3cm} \tess{} Quasi-Periodic GP Parameters$^c$:}  \\
$P_{\mathrm{GP}}$             &  GP kernel Period (days)                                  & $\mathcal{J}(0.001, 1000)$            \\ 
$B$                           &   Amplitude          & $\mathcal{J}(10^{-6},1)$      \\ 
$C$                           &  Constant scaling term                            & $\mathcal{J}(10^{-3},10^{3})$        \\ 
$L$  &  Characteristic time-scale &$\mathcal{J}(1, 10^{3})$          \\
\enddata
\tablenotetext{a}{We use the same uniform priors for pairs of limb darkening parameters $q_1$ and $q_2$ (parametrization from \cite{kipping2013}, and use separate limb darkening parameters for each instrument).}
\tablenotetext{b}{We placed a separate photometric jitter term and baseline offset term for each of the photometric instruments (\tess{}, Penn State Davey CDK 0.6 m and Perkin 17").}
\tablenotetext{c}{We do not use a GP for the two ground based photometric instruments (Penn State Davey CDK 0.6 m and Perkin 17") since the total observation duration for each was only about 4 hours.}
\end{deluxetable}

\begin{deluxetable*}{llc}
\tablecaption{Derived Parameters for the TOI-1728 System \label{tab:planetprop}}
\tablehead{\colhead{~~~Parameter} &
\colhead{Units} &
\colhead{Value}
}
\startdata
\sidehead{Orbital Parameters:}
~~~Orbital Period\dotfill & $P$ (days) \dotfill & $3.491510_{-0.000057}^{+0.000062}$\\
~~~Time of Periastron\dotfill & $T_P$ (BJD\textsubscript{TDB})\dotfill & $2458843.707_{-1.010}^{+0.697}$\\
~~~Eccentricity\dotfill & $e$ \dotfill & $0.057_{-0.039}^{+0.054}$\\
~~~Argument of Periastron\dotfill & $\omega$ (degrees) \dotfill & $45_{-187}^{+104}$\\
~~~Semi-amplitude Velocity\dotfill & $K$ (m/s)\dotfill &
$15.12_{-2.87}^{+3.04}$\\
~~~Systemic Velocity$^a$\dotfill & $\gamma$ (m/s)\dotfill & 
$1.86_{-1.91}^{+1.84}$\\
~~~RV trend\dotfill & $dv/dt$ (\unit{mm/s/day})   & $0.001\pm0.025$         \\ 
~~~RV jitter\dotfill & $\sigma_{HPF}$ (m/s)\dotfill & $0.46_{-0.44}^{+3.36}$\\
\sidehead{Transit Parameters:}
~~~Transit Midpoint \dotfill & $T_C$ (BJD\textsubscript{TDB})\dotfill & $2458843.27427\pm0.00043$\\
~~~Scaled Radius\dotfill & $R_{p}/R_{*}$ \dotfill & 
$0.074\pm0.002$\\
~~~Scaled Semi-major Axis\dotfill & $a/R_{*}$ \dotfill & $13.48\pm0.20$\\
~~~Orbital Inclination\dotfill & $i$ (degrees)\dotfill & $88.31_{-0.40}^{+0.58}$\\
~~~Impact Parameter\dotfill & $b$\dotfill & $0.39_{-0.15}^{+0.11}$\\
~~~Transit Duration\dotfill & $T_{14}$ (hours)\dotfill & $1.96\pm0.03$\\
~~~Photometric Jitter$^b$ \dotfill & $\sigma_{TESS}$ (ppm)\dotfill & $ 0.02_{-0.02}^{+1.96}$\\
~~~ & $\sigma_{Davey}$ (ppm)\dotfill & $2334.64_{-196.66}^{+197.77}$\\
~~~ & $\sigma_{Perkin}$ (ppm)\dotfill & $1062.40_{-272.43}^{+261.22}$\\
~~~Dilution$^c$\dotfill & $D_{TESS}$ \dotfill & $0.860\pm0.045$\\
\sidehead{Planetary Parameters:}
~~~Mass\dotfill & $M_{p}$ (\unit{M}$_\oplus$)\dotfill &  $26.78_{-5.13}^{+5.43}$\\
~~~Radius\dotfill & $R_{p}$  (\unit{R}$_\oplus$) \dotfill& $5.05_{-0.17}^{+0.16}$\\
~~~Density\dotfill & $\rho_{p}$ (g/\unit{cm^{3}})\dotfill & $1.14_{-0.24}^{+0.26}$\\
~~~Surface Gravity\dotfill & $g_{p}$ (cm/s$^2$)\dotfill & $1037_{-30}^{+29}$\\ 
~~~Semi-major Axis\dotfill & $a$ (AU) \dotfill & $0.0391\pm0.0009$\\
~~~Average Incident Flux\dotfill & $\langle F \rangle$ (\unit{10^5\ W/m^2})\dotfill & $0.785_{-0.035}^{+0.033}$ \\
~~~Planetary Insolation& $S$ (\unit{S}$_\oplus$)\dotfill & $57.78\pm3.48$ \\
~~~Equilibrium Temperature\(^{d}\)\dotfill & $T_{eq}$ (K)\dotfill & $767\pm8$\\
\enddata
\tablenotetext{a}{In addition to the Absolute RV from \autoref{tab:stellarparam}.}
\tablenotetext{b}{Jitter (per observation) added in quadrature to photometric instrument error.}
\tablenotetext{c}{Dilution due to presence of background stars in \tess{} aperture.}
\tablenotetext{d}{The planet is assumed to be a black body.}

\normalsize
\end{deluxetable*}

\subsection{Upper limit on Helium 10830 \AA~absorption}\label{sec:he10830}

\begin{figure}[!t] 
\centering
\includegraphics[width=0.48\textwidth]
{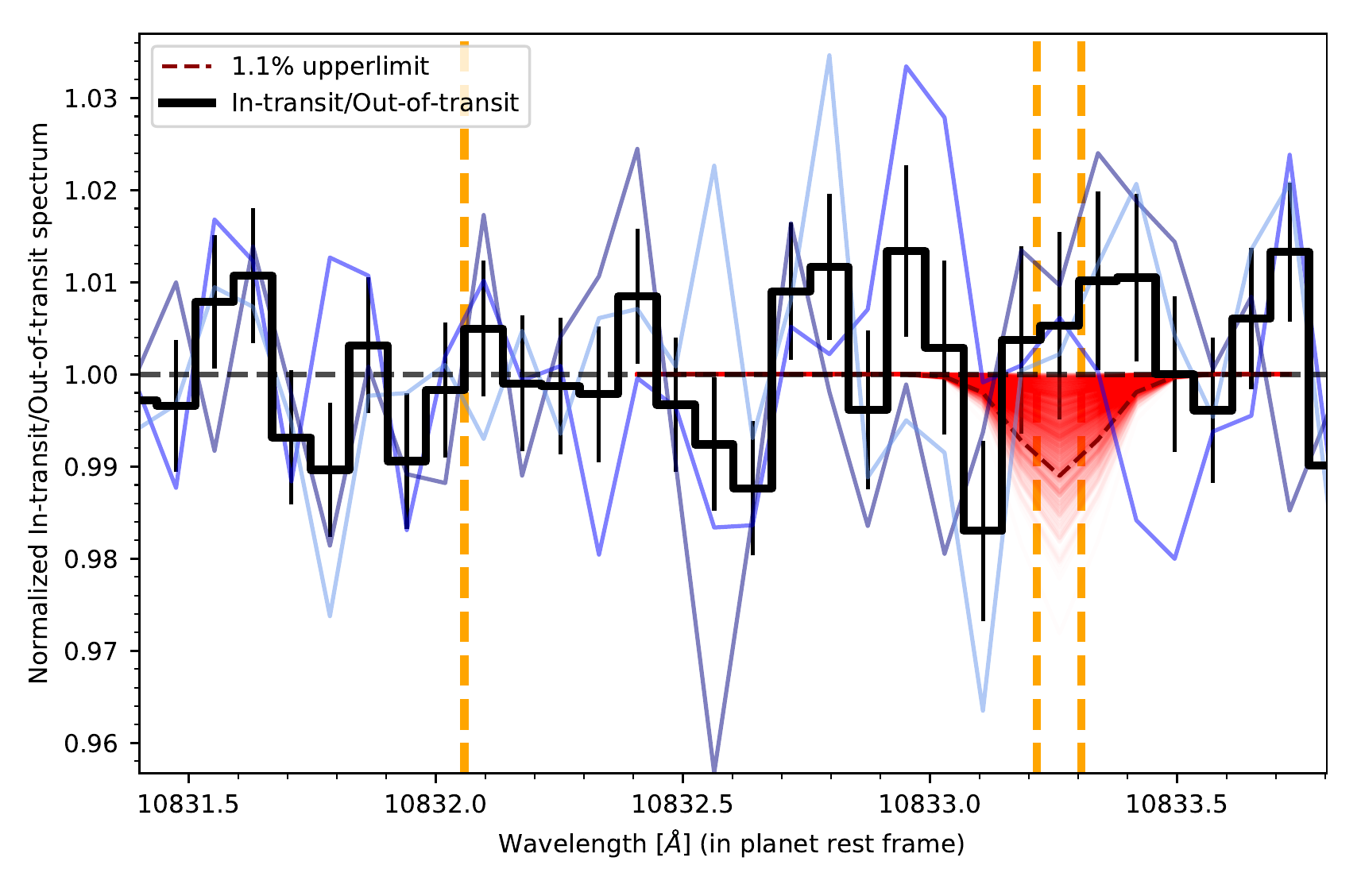} 
\caption{The ratio of the in-transit spectra and out-of-transit TOI-1728 spectra. The blue curves are the three individual ratio spectra from the transit epoch, whereas the black curve is the weighted average of the three. The x-axis shows vacuum wavelength in the planet's rest frame at mid-transit. The rest vacuum wavelengths of the He 10830 \AA, triplet lines in planet’s rest frame are marked by dashed vertical orange lines. We do not detect any significant absorption in the planetary spectra (lower panel) at these wavelengths. The results of our MCMC fit of the strongest doublet lines in the He 10830 \AA~triplet at the instrument resolution are shown by the red curves in the lower panel and the 1.1\% upper-limit is shown by the dashed red curve overlaid on the MCMC results.} \label{fig:he10830}
\end{figure}

He 10830 \AA  ~observations have recently emerged as a powerful ground-based probe to detect and constrain atmospheric outflow from hot Jupiters and warm Neptunes \citep{seager00,oklopcic18}. 
 The low bulk density of this planet makes it a promising candidate for large atmospheric mass outflows. We estimated an upper limit on the He 10830 \AA~signature from the spectrum obtained during the transit of TOI-1728. We observed a total of three spectra inside transit, with a median signal-to-noise ratio of $\sim$ 100 per pixel on 2020 March 1. A high S/N template of the out-of-transit spectrum was created by the averaging all the spectra we obtained for RV measurement outside the transit window. Careful subtraction of the sky emission line is crucial so as to not confuse sky lines with the He 10830 \AA~signal. We subtracted the simultaneous sky spectrum measured in the HPF's sky fiber after scaling by the throughput ratio obtained from twilight observations. \autoref{fig:he10830} shows the ratio of in-transit to the out-of-transit spectrum during the transit.  The error bars propagated through our pipeline are also shown in the plot. To calculate the upper limit we injected  artificial absorption lines and ran MCMC fits using  \texttt{PyMC3} \citep{exoplanet:pymc3} to tests detectability. The line width was taken to be HPF's instrument resolution, with the wavelength fixed to the two strong unresolved lines in the He 10830 \AA~ triplet, and the continuum normalized to 1. The posterior distribution of the amplitude of the absorption line was used to calculate its credible interval. With 90\% probability, we obtain the upper limit of the absorption trough to be $<$ 1.1\% (see Figure \ref{fig:he10830}). Deeper levels of He 10830 \AA ~ absorption have been detected in other planet atmospheres at the 90$\%$ confidence level using similar techniques by HPF and CARMENES \citep{Ninan2020, Palle2020}.

\section{Discussion}\label{sec:discussion}

\subsection{TOI-1728b in M dwarf planet parameter space}

\begin{figure*}
\gridline{\fig{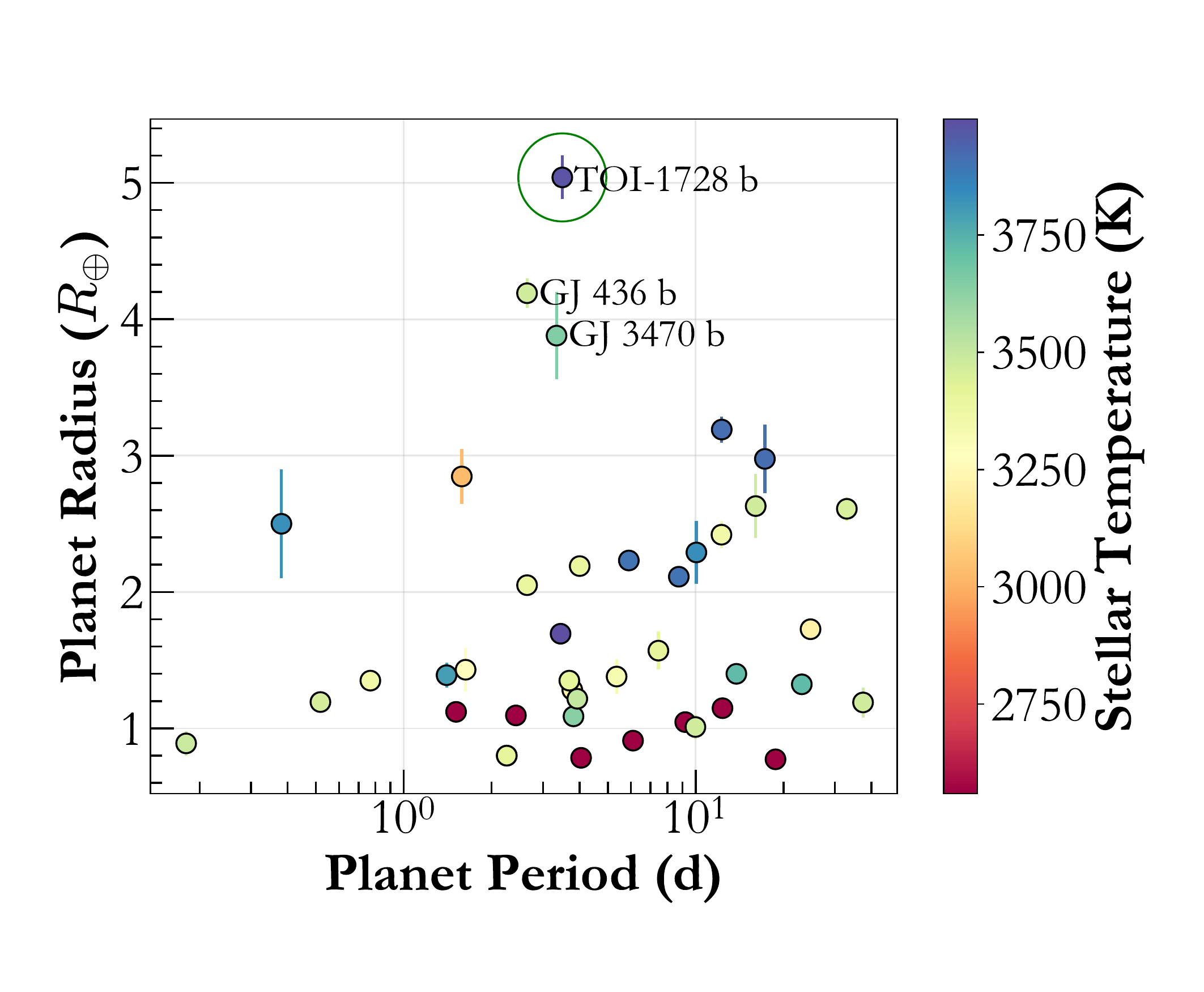}{0.45\textwidth}{{\small a) Period - Radius plane}}    \label{fig:RadiusPeriod}
          \fig{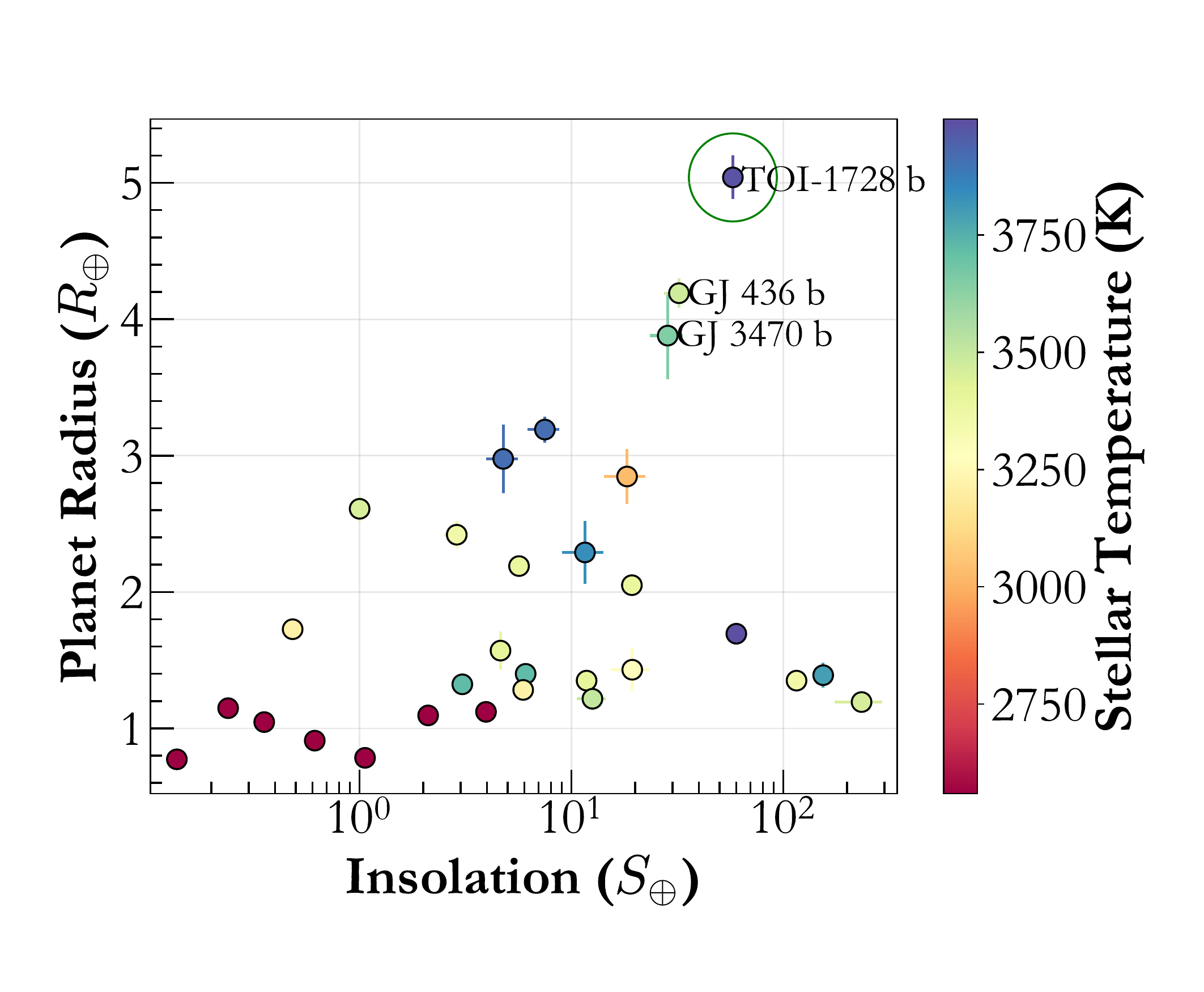}{0.45\textwidth}{ \small b) Insolation - Radius plane}} \label{fig:RadiusInsolation}
\gridline{\fig{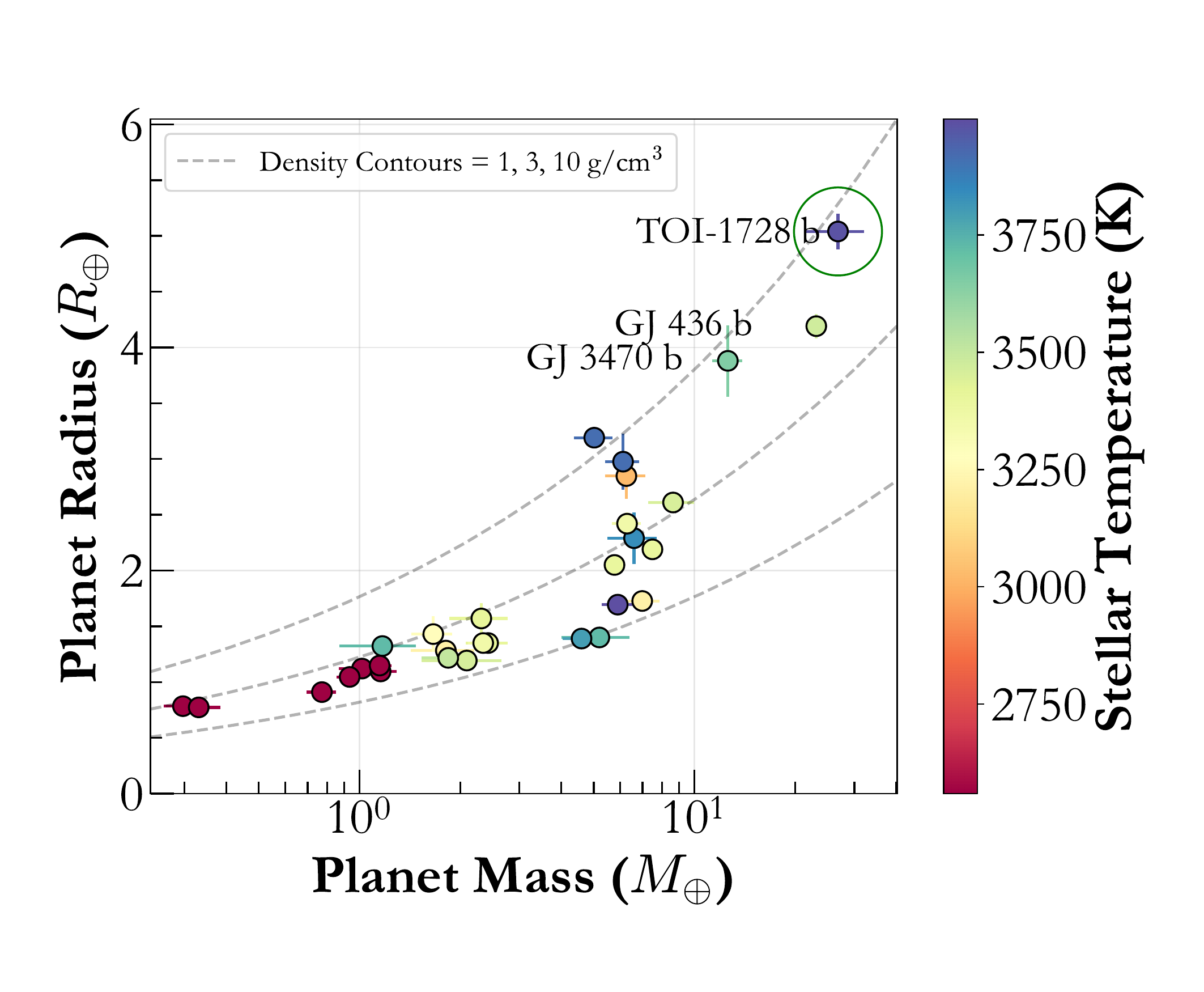}{0.45\textwidth}{\small c) Mass - Radius plane}   \label{fig:RadiusMass}
          \fig{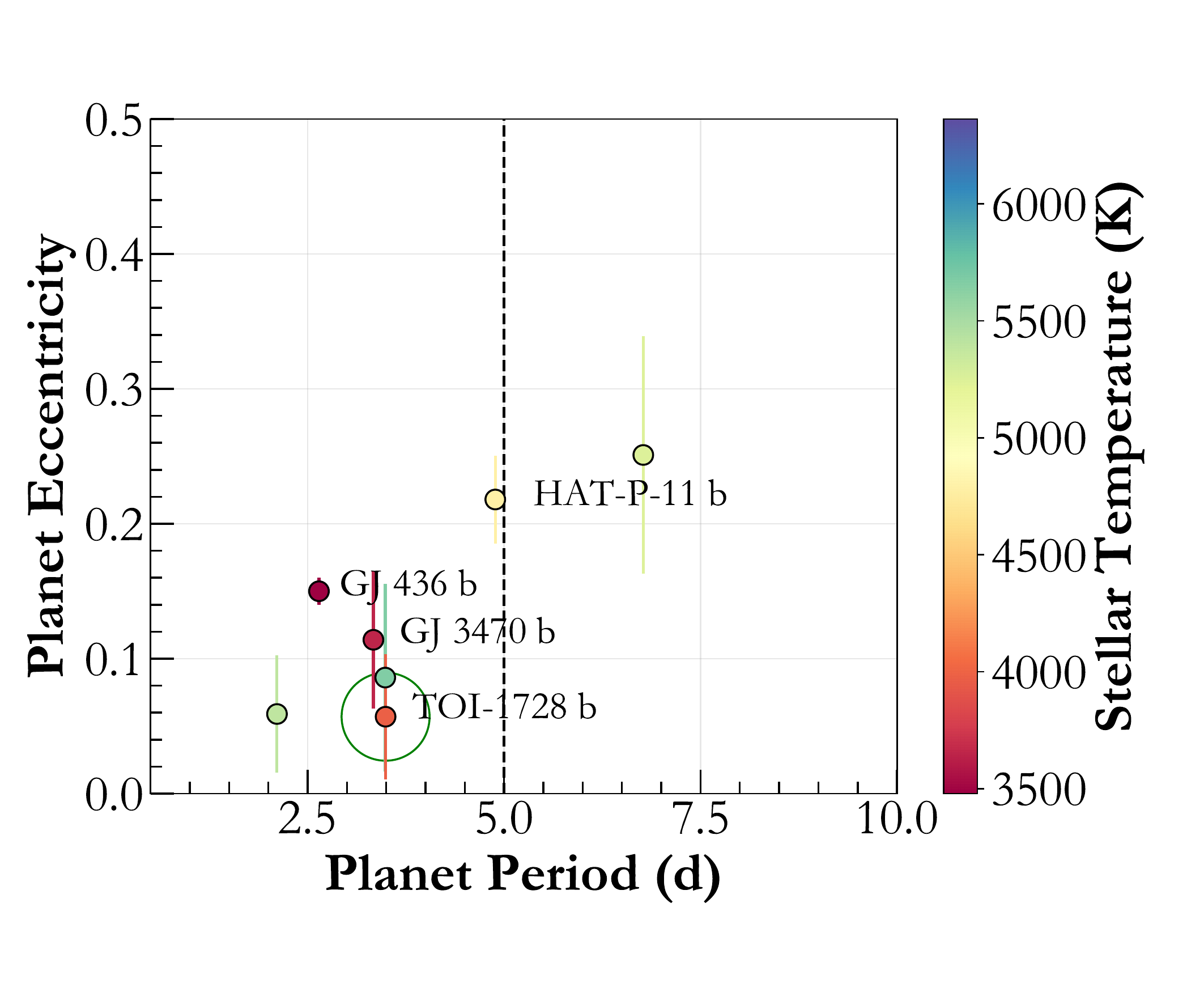}{0.45\textwidth}{\small d) Eccentricity - Period plane\label{fig:EccPeriod}}} 
\caption{\small We show TOI-1728b (circled) in different planet parameter space planes. \textbf{a, b, c)} Sample shown is limited to M dwarf planets ($\rm{R}_p < 6 ~\rm{R}_{\oplus}$). \textbf{d)} Neptune exoplanet sample ($3 \rm{R}_{\oplus} < \rm{R}_p < 6 \rm{R}_{\oplus}$) -- not limited to M dwarf exoplanets --  with eccentricity errors $<$ 0.1. \textbf{a)} Shows the Period-Radius plane, where TOI-1728b has comparable orbital periods to GJ 436b and GJ 3470b, but has a larger radius. \textbf{b)} TOI-1728b orbits an M0 host star which is is an earlier spectral type than the host for GJ 436b and GJ 3470b and hence receives higher insolation.  \textbf{c)} Mass - Radius plane for M dwarf planets with mass measurements at $> 3 \sigma$. We include contours of density 1, 3, 10 g/cm$^3$.   \textbf{d)} The general exo-Neptune population, where we highlight TOI-1728b. The vertical line marks the orbital period of 5 days, and is similar to the plot shown in \cite{correia_why_2020}. Our updated exoplanet sample includes TOI-1728b, for which the eccentricity is consistent with zero at 1$\sigma$.}\label{fig:PlanetParameters}
\end{figure*}

In \autoref{fig:PlanetParameters} we show where TOI-1728b lies in exoplanet parameter phase space compared to other known exoplanets.  For the purposes of illustration, we draw our sample from the NASA Exoplanet Archive \citep{akeson_nasa_2013} and further include recent transiting planets discovered by \tess{} around M dwarf stellar hosts, we include only those planets with either mass measurements or upper limits. For \autoref{fig:PlanetParameters} a, b, c we restrict our sample to \teff{} $<$ 4000 K and planetary radii  $R_p < 6 R_{\oplus}$, whereas for \autoref{fig:PlanetParameters} c we impose an additional requirement of a 3$\sigma$ mass measurement. For \autoref{fig:PlanetParameters} d, we restrict our sample to planets with \teff{} $<$ 7000 K,  planetary radius - $3 \rm{R}_{\oplus} < \rm{R}_p < 6 \rm{R}_{\oplus}$, and eccentricity errors $<$ 0.1. We also looked at the exoplanet sample for host stars of mass up to 0.75 M${\odot}$, however this did not add any super Neptunes comparable to TOI-1728b to the discussion, and hence we limit our planet sample to exoplanets around M dwarfs with \teff{} $<$ 4000 K.

TOI-1728b (M0 host star) has an orbital period comparable to GJ 3470b and GJ 436b (Figure \ref{fig:PlanetParameters}a), which orbit a M1.5V and M2.5V host star, respectively. Due to the hotter effective temperature, TOI-1728b receives higher insolation flux than the other two comparable warm Neptunes (Figure \ref{fig:PlanetParameters}b). TOI-1728b also represents an important addition to the Mass-Radius plane (Figure \ref{fig:PlanetParameters}c) for M-dwarf exoplanets. The current sample of transiting M dwarf exoplanets consists of about 40 planets with mass measurements (and not only upper limits). Of these, the only other comparable planets are GJ 3470b and GJ 436b. There have been numerous studies which find differences in the occurrence rates for planets around M dwarfs vs earlier type stars \citep{2019AJ....158...75H, 2020arXiv200202573H}, as well as the empirical distributions of mass and radius for said planets \citep{bonfils_harps_2013, dressing_occurrence_2015}. These results motivate independent statistical studies of M dwarf exoplanet populations \citep{Kanodia2019, cloutier_evolution_2019}. However, the aforementioned studies suffer from a small M dwarf planet sample. Furthermore, the large range of stellar masses in the M dwarf spectral type (0.65 M$_{\odot}$ to 0.08  M$_{\odot}$) means that we can possibly expect differences in planetary formation and properties within the spectral subtype due to variation in luminosity, protoplanetary disk mass and composition.  We need a larger sample of M dwarf exoplanets to study these trends in planetary formation and evolution, for which TOI-1728b represents a step forward.

Using existing compositional models \citep{2014ApJ...792....1L} we estimate TOI-1728b to have a H/He atmospheric envelope of 15 -- 25$\%$ mass fraction (H/He + rock). We get comparable estimates from the \cite{2008A&A...482..315B} models which predict a H/He mass fraction $> 10\%$.
 However, we would advise caution on this estimate, because these models  \citep{2007ApJ...659.1661F, 2014ApJ...792....1L, 2019PNAS..116.9723Z} use an exoplanet sample which consists mostly of solar type stellar hosts, while there have been studies which suggest that the planetary formation mechanism has a dependence on stellar mass \citep{mulders_stellar-mass-dependent_2015, mulders_increase_2015}.  Also, most of these models are generally limited to rocky planets and sub Neptunes with R$_p < $ 4 -- 4.5 R$_{\oplus}$.  The H/He envelope of Neptunes orbiting solar type stars tend to have a higher metallicity [M/H] than Jupiter sized planets due to planetesimal accretion \citep{2016A&A...596A..90V}. However it is not clear whether this should be seen in Neptunes orbiting M dwarfs as well, due to different timescales for planetesimal accretion and disk lifetime for M dwarf hosts vs.\ solar type host stars \citep{Ogihara_2009}.

\subsection{Eccentricity of TOI-1728b}
The characteristic circularization time scale for most  warm Neptunes is typically $< 5$ Gyr, and hence most observed Neptunes should be in circular orbits. However, most warm Neptunes (P $<$ 5 days) tend to exhibit non zero eccentricity at $> 1 \sigma$ \citep[][]{correia_why_2020};  and GJ 436b \citep{turner_ground-based_2016}, HAT-P-11b  \citep{2018AJ....155..255Y}, are all eccentric at $> 3 \sigma$ (Figure \ref{fig:PlanetParameters}d). This eccentricity distribution of warm Neptunes contrasts with that for hot Jupiters and rocky planets, where the shorter orbital period planets tend to have circular orbits due to tidal dissipation. \cite{correia_why_2020} discuss multiple mechanisms which oppose bodily tides to slow down this circularization process; a combination of which (thermal tides in the atmosphere, atmospheric escape, or excitation due to a companion) can potentially yield the observed population of eccentric Neptunes.

TOI-1728b, is a warm super Neptune with period $\sim$ 3.5 days and eccentricity $0.057_{-0.039}^{+0.054}$ (Figure \ref{fig:PlanetParameters}). The current eccentricity estimate for TOI-1728b is consistent with both a circular and mildly eccentric orbit. While it does rule out a highly eccentric system, it does not have the precision for more substantial claims regarding its place in the eccentric warm Neptune population. This constraint could be greatly improved with additional observations, especially the observation of a secondary eclipse. We also recognize that given the truncated positive semi-definite nature of the eccentricity distribution, there tends to exist a bias towards higher values in eccentricity estimates \citep{1971AJ.....76..544L, 2008ApJ...685..553S}. Considering these factors, if TOI-1728b is indeed in a circular orbit, this would contrast with the eccentricity distribution for warm Neptunes seen by \cite{correia_why_2020}, and could be attributed to a few possible hypotheses:
\begin{itemize}
    \item The TOI-1728 stellar system is old enough to circularize the planet despite the competing mechanisms. From Equation 2 of \cite{correia_why_2020}, the characteristic circularization time scale for TOI-1728b is $\sim$ 0.8 Gyr, while our age estimates from the SED fit predict a stellar age of 7.0 $\pm$ 4.6 Gyr. This estimate, coupled with the lack of a detectable rotation period in photometry or RVs, and the lack of stellar activity indicate an old stellar host system which has had time to circularize its orbit.
    \item The initial orbit of TOI-1728b was not as highly eccentric as the other warm Neptunes in the \cite{correia_why_2020} sample. Since the total time for circularization scales with the initial eccentricity, if TOI-1728b formed in an orbit which was not highly eccentric, then it would be easier to circularize it. 
    \item A companion object can pump up the eccentricity of a planet by the  Kozai-Lidov mechanism \citep{1962AJ.....67..591K, 1962P&SS....9..719L} or Spin-Eccentricity pumping \citep{2013A&A...553A..39C, 2013CeMDA.117..331G}. For TOI-1728b, the \tess{} light curve does not show another transiting companion, however this does not rule out non-transiting companions or those with orbital periods $\gtrsim$ 27 days. Furthermore, the RV residuals do not show a periodic signal, which at least rules out a massive and short period companion object. 
\end{itemize}

\subsection{Atmospheric Escape}\label{sec:atmescape}
Based on energy conservation, in a planet heated by host star's irradiation, the atmosphere mass escape rate is proportional to the XUV\footnote{XUV includes  X-ray and extreme ultraviolet photons ($<$ 912~\AA).} flux falling on the planet, and inversely proportional to the density of the planet \citep{SanzForcada2011}. 
We summarize the detections for He 10830 \AA ~ and Ly$\alpha$ absorption for GJ436b, GJ3470b --- both warm Neptunes around M dwarfs --- as well as HAT-P-11b, a warm Neptune around a mid K dwarf, as they represent the closest analogues to TOI-1728b with intensive transmission spectroscopy follow-up.

\textbf{He 10830 \AA:} Transit spectroscopy measurements have detected absorption in the NIR corresponding to ionized He 10830 \AA ~ for GJ3470b \citep{Ninan2020,Palle2020}, and also HAT-P-11b \citep[][]{2018AJ....155..255Y, 2018ApJ...868L..34M, 2018Sci...362.1384A}. At the same time, no absorption corresponding to this feature has been detected for GJ436b \citep{Nortmann_2018} or for TOI-1728b (this work).

\textbf{Ly$\alpha$ measurements: } Ly$\alpha$ observations using Hubble Space Telescope (HST) have detected significant atmospheric mass outflows in GJ 3470b \citep{bourrier18} and GJ 436b \citep{kulow14, ehrenheich_2015}.

GJ 436b represents a particularly interesting case with Ly$\alpha$ detection in the UV, but lack of He 10830 \AA ~ absorption.  This could be due to a lower helium ionizing flux in comparison to the hydrogen ionizing flux, and even to certain extent a fractional under-abundance of Helium in the exo-sphere of GJ 436b \citep{Hu_2015}.

Similarly for TOI-1728b, despite having lower planet density, and higher stellar irradiance than GJ3470b, our upper-limit of 1.1\% in He 10830 \AA~ absorption for TOI-1728b is less than the 1.5\% absorption detection seen in GJ 3470b. This could imply a lesser helium ionising flux from TOI-1728 than GJ 3470, and a scenario similar to that for GJ 436b. Given the similarity in planetary properties for these planets, we encourage the follow-up of TOI-1728 using the HST for Ly$\alpha$ exosphere detection, as well as further NIR observations to put tighter constraints on the upper limit  on He 10830 \AA ~ absorption.

\subsection{Potential for Transmission spectroscopy}

In addition to the atmospheric escape transmission spectroscopy observations detailed in the previous section, TOI-1728b is also a promising candidate for atmospheric abundance characterization.

\begin{figure}[!t] 
\centering
\includegraphics[width=0.5\textwidth]{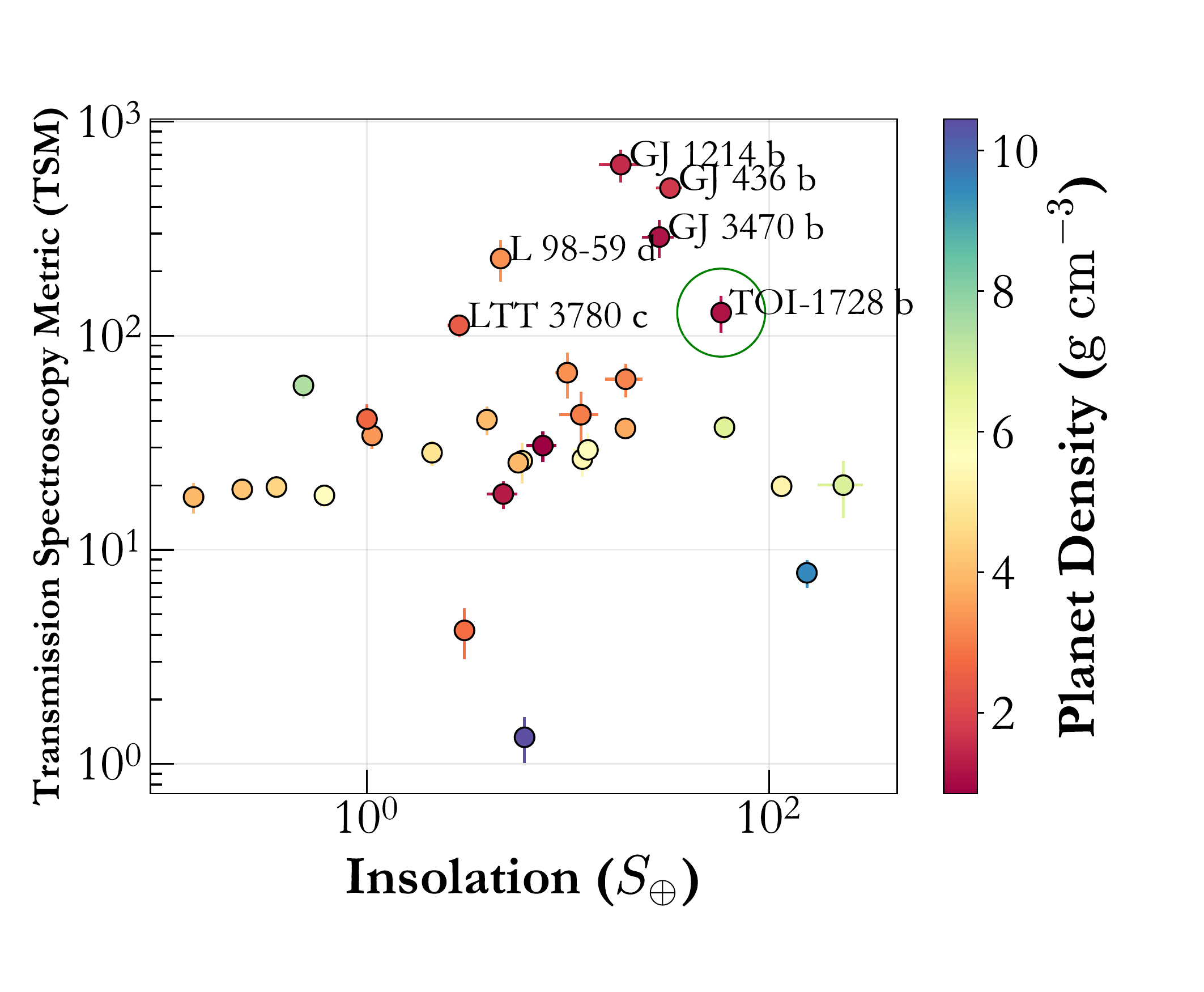}
\caption{TSM plot showing TOI-1728b with respect to other M dwarf planets with mass measurements  \citep{Kempton2018}. TOI-1728b's low planetary density and relatively warm equilibrium temperature contributes towards a high TSM.} \label{fig:tsm}
\end{figure}

 Warm Neptunes (800 - 1200 K)  represent equilibrium temperatures where the chemical and dynamical timescales become comparable and the assumption of chemical equilibrium breaks down \citep{2020arXiv200410106G}. An example of this is GJ 436b, where transmission spectroscopy in the infrared with \textit{Spitzer} has revealed that the relative abundances of CO to CH$_4$ are higher than expectations \citep{2011ApJ...729...41M}.  \cite{Crossfield2017} calculate the correlations between the detectability of H$_2$O signal for Neptunes with the equilibrium temperature and the H/He. These, when applied to TOI-1728b indicate a potential scale height for water of 1-2 (in units of H/He scale height), which would make TOI-1728b a good target for atmospheric characterization.
 
The Transmission Spectroscopy Metric \citep[TSM,][]{Kempton2018} for TOI-1728b is $\sim$ 130, which  is the 5$^{\rm{th}}$ highest TSM of sub-Jovian M dwarf planets ($\rm{R}_p < 9 ~\rm{R}_{\oplus}$) with mass measurements. We limit our TSM plot (\autoref{fig:tsm}) to planets with mass measurements (3$\sigma$ or better), because mass measurements are required to place  priors on the atmospheric scale height,  which is used to estimate the S/N of transmission spectroscopy observations. 
TOI-1728b has a density of $1.14_{-0.24}^{+0.26}$ g/cm$^3$ and an inflated nature that lends itself to be particularly suitable for transmission spectroscopy. Our mass measurement precision of 5.1$\sigma$ is essential in order to derive posteriors for atmospheric features that are not limited by the uncertainties in mass \citep{batalha_precision_2019}.

\section{Summary}\label{sec:summary}
In this work, we report the discovery and confirmation of a super-Neptune, TOI-1728b, orbiting an old and inactive M0 star in a $\sim$ 3.5 day circular orbit.  We detail the \tess{} photometry, as well as the ground-based follow-up photometry and RV observations using HPF. We also observe the planet in transit and claim a null detection of He 10830 \AA~ absorption with an upper limit of 1.1$\%$ which points to a inactive star with low X-ray emissions. For comparison, deeper levels of He 10830 \AA ~absorption have been detected in comparable Neptunes. It has an eccentricity consistent with both a circular and mildly eccentric orbit, and would benefit from more RV observations as well as a secondary eclipse to obtain a tighter eccentricity constraint. If indeed circular, this would represent an interesting departure from observed trends in the eccentricity of warm Neptunes. With a density of $1.14_{-0.24}^{+0.26}$ g/cm$^3$ TOI-1728b represents a inflated gaseous planet with the fifth highest TSM among sub Jovian M dwarf planets with mass measurements. In combination with the predicted scale height for water, this high TSM makes it a great target for follow-up with transmission spectroscopy with HST and JWST. TOI-1728b is the largest transiting super-Neptune around an M dwarf host, and characterization of its atmosphere can help better inform theories of planetary evolution and formation. Therefore we encourage future transmission spectroscopy observations of this target to characterize its atmosphere, as well as measure atmospheric escape.

\section*{Acknowledgement}
The authors thank the referee for providing insightful comments, which improved this work and made the results clearer.
This work was partially supported by funding from the Center for Exoplanets and Habitable Worlds. The Center for Exoplanets and Habitable Worlds is supported by the Pennsylvania State University, the Eberly College of Science, and the Pennsylvania Space Grant Consortium.
These results are based on observations obtained with the Habitable-zone Planet Finder Spectrograph on the HET. We acknowledge support from NSF grants AST 1006676, AST 1126413, AST 1310875, AST 1310885, and the NASA Astrobiology Institute (NNA09DA76A) in our pursuit of precision radial velocities in the NIR. We acknowledge support from the Heising-Simons Foundation via grant 2017-0494.  The Hobby-Eberly Telescope is a joint project of the University of Texas at Austin, the Pennsylvania State University, Ludwig-Maximilians-Universität München, and Georg-August Universität Gottingen. The HET is named in honor of its principal benefactors, William P. Hobby and Robert E. Eberly. The HET collaboration acknowledges the support and resources from the Texas Advanced Computing Center. We thank the Resident astronomers and Telescope Operators at the HET for the skillful execution of our observations of our observations with HPF. 
We acknowledge support from NSF grant AST-1909506 and the Research Corporation for precision photometric observations with diffuser-assisted photometry.

This research has made use of the NASA Exoplanet Archive, which is operated by the California Institute of Technology, under contract with the National Aeronautics and Space Administration under the Exoplanet Exploration Program. 
This work includes data collected by the \tess{} mission, which are publicly available from MAST. Funding for the \tess{} mission is provided by the NASA Science Mission directorate. 
Some of the data presented in this paper were obtained from MAST. Support for MAST for non-HST data is provided by the NASA Office of Space Science via grant NNX09AF08G and by other grants and contracts.
This research has made use of NASA's Astrophysics Data System Bibliographic Services.

Part of this research was carried out at the Jet Propulsion Laboratory, California Institute of Technology, under a contract with the National Aeronautics and Space Administration (NASA).
Computations for this research were performed on the Pennsylvania State University’s Institute for Computational and Data Sciences Advanced CyberInfrastructure (ICDS-ACI), including the CyberLAMP cluster supported by NSF grant MRI-1626251.
This work includes data from 2MASS, which is a joint project of the University of Massachusetts and IPAC at Caltech funded by NASA and the NSF.
We acknowledge with thanks the variable star observations from the AAVSO International Database contributed by observers worldwide and used in this research.
This work has made use of data from the European Space Agency (ESA) mission \gaia{} (\url{https://www.cosmos.esa.int/gaia}), processed by the \gaia{} Data Processing and Analysis Consortium (DPAC, \url{https://www.cosmos.esa.int/web/gaia/dpac/consortium}). Funding for the DPAC has been provided by national institutions, in particular the institutions participating in the \gaia{} Multilateral Agreement.
Some observations were obtained with the Samuel Oschin 48-inch Telescope at the Palomar Observatory as part of the ZTF project. ZTF is supported by the NSF under Grant No. AST-1440341 and a collaboration including Caltech, IPAC, the Weizmann Institute for Science, the Oskar Klein Center at Stockholm University, the University of Maryland, the University of Washington, Deutsches Elektronen-Synchrotron and Humboldt University, Los Alamos National Laboratories, the TANGO Consortium of Taiwan, the University of Wisconsin at Milwaukee, and Lawrence Berkeley National Laboratories. Operations are conducted by COO, IPAC, and UW.
We gathered data using the Planewave CDK 24 Telescope operated by the Penn State Department of Astronomy $\&$ Astrophysics at Davey Lab Observatory.

SK would like to acknowledge useful discussions with Dan Foreman-Mackey regarding the \texttt{exoplanet} package and Theodora for help with this project. CIC acknowledges support by NASA Headquarters under the NASA Earth and Space Science Fellowship Program through grant 80NSSC18K1114.

This research made use of \textsf{exoplanet} \citep{exoplanet:exoplanet} and its dependencies \citep{exoplanet:agol19, exoplanet:astropy13, AstropyCollaboration2018, Foreman-Mackey2017, exoplanet:foremanmackey18,exoplanet:kipping13, exoplanet:luger18, exoplanet:pymc3, exoplanet:theano}.

\facilities{\gaia{}, HET (HPF), \tess{}, Perkin 17", Penn State CDK0.6 m, Exoplanet Archive}
\software{AstroImageJ \citep{Collins2017}, 
\texttt{astroquery} \citep{Ginsburg2019}, 
\texttt{astropy} \citep{exoplanet:astropy13, AstropyCollaboration2018},
\texttt{barycorrpy} \citep{Kanodia2018}, 
\texttt{batman} \citep{Kreidberg2015},
\texttt{celerite} \citep{Foreman-Mackey2017},
\texttt{corner.py} \citep{corner},
\texttt{dustmaps} \citep{dustmaps},
\texttt{dynesty} \citep{Speagle2019},
\texttt{EXOFASTv2} \citep{Eastman2019},
\texttt{exoplanet} \citep{exoplanet:exoplanet},
\texttt{HxRGproc} \citep{Ninan2018},
\texttt{GALPY} \citep{2015ApJS..216...29B}, 
\texttt{GNU Parallel} \citep{Tange2011},
\texttt{juliet} \citep{Espinoza2019},
 \texttt{lightkurve} \citep{LightkurveCollaboration2018},
\texttt{matplotlib} \citep{hunter2007},
\texttt{MRExo} \citep{Kanodia2019},
\texttt{numpy} \citep{vanderwalt2011},
\texttt{pandas} \citep{McKinney2010},
\texttt{Photutils} \citep{Bradley2019},
\texttt{PyMC3}\citep{exoplanet:pymc3},
\texttt{radvel}\citep{Fulton2018},
\texttt{scipy} \citep{Virtanen2019},
\texttt{SERVAL} \citep{Zechmeister2018},
\texttt{starry}\citep{exoplanet:luger18},
\texttt{Theano}\citep{exoplanet:theano}
}

\bibliography{MyLibrary}

\listofchanges
\end{document}